\documentclass{article}

\usepackage{arxiv}
\usepackage{appendix}
\usepackage[utf8]{inputenc}
\usepackage[T1]{fontenc}   
\usepackage{booktabs}      
\usepackage{amsfonts}      
\usepackage{nicefrac}      
\usepackage{microtype}     
\usepackage{lipsum}		
\usepackage{graphicx}
\usepackage{doi}
\usepackage{graphicx}
\usepackage{appendix}
\usepackage{derivative}
\usepackage{amsmath}
\usepackage{amssymb}
\usepackage{wrapfig}

\usepackage{dcolumn}
\usepackage{bm}
\usepackage[hang]{subfigure}
\usepackage[section]{placeins}
\usepackage{siunitx}
\usepackage{float}
\usepackage{multirow}

\usepackage[symbol]{footmisc}

\usepackage{hyperref}

\makeatletter
\def\hlinewd#1{
\noalign{\ifnum0=`}\fi\hrule \@height #1 \futurelet
\reserved@a\@xhline}
\makeatother

\title{Anticipation of Oligocene's climate heartbeat by simplified eigenvalue estimation}

\author{\href{https://orcid.org/0000-0003-0529-7926}{\includegraphics[scale=0.06]{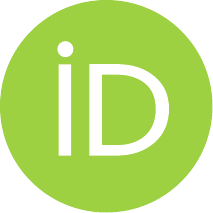}\hspace{1mm}Martin ~Heßler}\thanks{Center for Nonlinear Science, University of Münster, 48149 Münster, Germany} \\
	Institute for Theoretical Physics\\
	University of Münster\\
	48149 Münster, North Rhine-Westphalia, Germany\\
	\texttt{m\_{}hess23@uni-muenster.de} \\
    \And
	\href{https://orcid.org/0000-0003-0986-0878}{\includegraphics[scale=0.06]{orcid.pdf}\hspace{1mm}Oliver ~Kamps} \\
	Center for Nonlinear Science\\
	University of Münster\\ 
	48149 Münster, North Rhine-Westphalia, Germany\\
	\texttt{okamp@uni-muenster.de} \\
}

\renewcommand{\headeright}{\emph{arXiv} Preprint}
\renewcommand{\undertitle}{\emph{arXiv} Preprint}
\renewcommand{\shorttitle}{Anticipation of Oligocene's climate heartbeat}

\hypersetup{
pdftitle={Anticipation of Oligocene's climate heartbeat by simplified eigenvalue estimation},
pdfsubject={physics.data-an nlin.CD},
pdfauthor={Martin Heßler, Oliver Kamps},
pdfkeywords={Eocene-Oligocene climate transition, dominant eigenvalue estimation, Neimark-Sacker bifurcation, Hopf bifurcation, bifurcation type classificaiton, limit cycle stability analysis ,early warning signal, leading indicator},
}

\begin{document}
\maketitle

\begin{abstract}
The Eocene-Oligocene transition marks a watershed point of earth's climate history. The climate shifts from a greenhouse state to an icehouse state in which Antarctica glaciated for the first time and periodic dynamics arise which are still relevant for our current climate. We analyse a $CaCO_3$ concentration time series which covers the Eocene-Oligocene transition and which is obtained from a Pacific sediment core at site DSDP1218. Therefore, we introduce a simplified autoregression-based variant of the dominant eigenvalue (DEV) estimation procedure. The DEV works as leading indicator of bifurcation-induced transitions and enables us to identify the bifurcation type. We confirm its reliability in a methodological study and demonstrate the crucial importance of proper detrending to obtain unbiased results. As a remark, we discuss also possible pathways to estimate the stability of limit cycles based on the DEV and the alternative drift slope as a proof of principle. Finally, we present the DEV analysis results of the $CaCO_3$ concentration time series which are reproducible in a wide parameter range. Our findings demonstrate that the onset of Oligocene's periodic dynamics might be announced by a Neimark-Sacker/Hopf bifurcation in course of the Eocene-Oligocene transition $\SI{34}{mya}$\footnote[4]{We follow the convention and use mya $\widehat{=}$ ``million years ago'' and Ma $\widehat{=}$ ``million years'' throughout the article.}.

\end{abstract}


\section{\label{sec:introduction}Introduction}
The Eocene-Oligocene climate transition~\cite{a:Tripati2005} from a warm greenhouse climate state to a colder icehouse state with glaciated polar regions $\SI{34}{mya}$ is of special interest for today's climate research: The mentioned transition marks the earliest beginning of the current climate state and is accompanied by the first glaciation of Antarctica~\cite{a:Tripati2005,a:Galeotti2016}. Furthermore, the Oligocene manifests similar dynamics as our current Quaternary~\cite{ic:Ehlers2011}. Keeping that in mind, a better understanding of the transition dynamics is crucial for model improvement and interpretation of the current climate state.\\
Paelike et al.~\cite{a:Paelike2006} investigated data from a Pacific sediment core that covers $\SI{13}{Ma}$ of the Oligocene without interruption. They observed first the ``Oligocene's climate heartbeat'' that gave the name to our article. This heartbeat is characterized by pronounced oscillations of glacial-interglacial periods and of the carbon cycle. The exact mechanisms of that heartbeat are not fully understood, but the authors show that a Milankovitch-type~\cite{b:milankovitch1920} astronomical forcing of the carbon cycle, the atmospheric $CO_2$ concentration, the temperature and glaciations play an important role. Similar observations have been made for the current Quaternary period that began $\SI{2.6}{mya}$: Periodic glacial-interglacial cycles~\cite{ic:Maslin,a:Boers2022} with a period length of $\SI{100}{ka}$ are observed for the last $\SI{1}{Ma}$. The time interval's climate is also subject to periodic variations in atmospheric gas concentrations, e.g. $CO_2$ and methane $CH_4$~\cite{ic:Ehlers2011,ic:Stauffer}. Even if climate scientists mostly agree with the Milankovitch forcing of the glacial-interglacial cycle, understanding the relative role of intrinsic nonlinear aspects of Earth's climate system and the above mentioned external forcing gives rise to ongoing debate~\cite{a:Boers2022, a:ghil1984, a:Huybers2009, a:Riechers2022}. In contrast, intrinsic nonlinear oscillation dynamics without feedback are supposed to be responsible for the Dansgaard-Oeschger events~\cite{a:Peltier2014,a:Dokken2013}. All these periodic phenomena are not known for the Eocene climate~\cite{a:Evans2018}, although the Quaternary, Oligocene and the Eocene are periods of the current Cenozoic era~\cite{a:stratigraphy}.\\
In this study we focus on the actual transition from the warm Eocene climate to the Oligocene climate with its periodic features. The mechanisms of the transition are still subject to debate~\cite{a:Boers2022}. Changes in the $CO_2$ concentration, Milankovitch forcing and modified oceanic gateways and currents (e.g. Tasman and Neotethys gateway~\cite{a:Kennedy2015,a:Jovane2009}, the Greenland–Scotland Ridge and the proto–Fram Strait~\cite{a:Straume2022}, Ross sea gyre circulation~\cite{a:Kennedy2015})  might have triggered the transition all together or partially with variable impact~\cite{a:Boers2022,a:Ladant2014}. However, the mathematical nature of the transition remains an open research task~\cite{a:Boers2022}.\\
Not only in the context of sudden abrupt climate transitions the research topic of tipping points and leading indicators becomes more important~\cite{a:Kaszas2019}. Commonly known mechanisms as bifurcation-induced (B-tipping), rate-dependent (R-tipping) and noise-induced tipping (N-tipping) are discussed in~\cite{a:Ashwin2012,a:Ritchie2017,a:Feudel2023}. Briefly summarized, B-tipping denotes a transition due to a control parameter exceeding a critical threshold which leads to a bifurcation. R-tipping is due to dynamics on multiple time scales and particularly fast control parameter shifts even without crossing a critical threshold. N-tipping is e.g. a system jumping between two alternative stable states due to strong or accumulating noise contributions which kick the system over the stable basin's boundary. But also more exotic transitions between different chaotic attractors or fractality-induced tipping between two qualitatively different regular attractors and much more can be found in the literature~\cite{a:Kaszas2019}. Various climate subsystems which might manifest probably tipping risks are identified~\cite{a:Lenton2008}. In some cases, tipping of a subsystem might trigger further subsystems to tip in a tipping cascade~\cite{a:Klose2021}. However, most of the identified systems at risk have not yet been shown to exhibit unambiguous signs of tipping events~\cite{a:Lenton2008}. The same holds for the Eocene-Oligocene climate transition~\cite{a:Boers2022,a:Boettner2021}.\\
In the recent literature, the anticipation of B-tipping is a rather appealing scientific research idea. Based on the mathematical theory of complex systems, time series of various fields of research have been investigated with more or less success: e.g. the oxygen concentration in the atmosphere~\cite{a:livina15}, the onset of epileptic seizures (but without success~\cite{a:Wilkat2019}), narcolepsy~\cite{a:rikkert16, a:izrailtyan} or depression~\cite{a:leemput13}, the instability of ecological systems~\cite{a:dose06, a:gsell16, a:Hessler2022}, the outbreak of war~\cite{a:chadefaux14} (with conflict-related news data) and much more~\cite{a:scheffer12}. Some of them seem to share generic mathematical features that allow for an anticipation of a critical transition that can cause completely different behaviour than observed in the pre-transition state. These generic and common features are thought to be a consequence of so-called \textit{critical slowing down} prior to a bifurcation. Critical slowing down means that the relaxation rate to the stable equilibrium state decreases. That can be imagined by the flattening of the system's stable potential valley due to a control parameter approaching a critical threshold. This decreased relaxation rate corresponds to more similar values of subsequent equidistantly sampled time series points, thus an increased autocorrelation and a higher variance at the \textit{same} time~\cite{a:Ditlevsen2010, a:scheffer12, a:dakos12}.\\
Since the $CaCO_3$ climate proxy time series~\cite{d:Tripati2005} gained from a Pacific sediment core at site DSDP1218 covers the Eocene-Oligocene transition, it was already subject to several studies~\cite{a:Tripati2005, a:dakos08, a:grziwotz} and will be reconsidered in our article. Dakos et al.~\cite{a:dakos08} report an increase in autocorrelation prior to the transition. Unfortunately, this cannot be seen as a trustworthy sign of critical slowing down and B-tipping, because the variance exhibits a negative trend (cf. Appendix \ref{subsec: appendix std greenhouse}) which contradicts the claim of simultaneous increase of variance and autocorrelation. Apart from these so-called statistical leading indicators the dominant eigenvalue (DEV) of the Jacobian of the linearised system also corresponds to a decreased relaxation time. Furthermore, it provides the opportunity to classify the bifurcation types by the DEV trend in the Gaussian plane~\cite{a:grziwotz}. In particular, the method is able to distinguish flip and Hopf bifurcations from other bifurcation types, as e.g. fold bifurcations. A slightly better hint to the hypothesis of B-tipping underlying the Eocene-Oligocene transition is the positive absolute DEV trend found by Grziwotz et al.~\cite{a:grziwotz}. However, the DEV estimates in the Gaussian plane are not unambiguous for the claimed fold bifurcation. Instead of being purely real as they are supposed to be, they are complex with small imaginary parts.\\ 
In this article we derive a simpler implementation of the method introduced by Grziwotz et al.~\cite{a:grziwotz}, replacing the original parameter estimation with sequential maps (S-maps)~\cite{ip:deyle} by an autoregression scheme of order $p$ (AR($p$)). We provide an alternative procedure to find an optimal autoregression order $p$ by a false next neighbour (fnn) classification and perform a methodological study on this simplified method applied to synthetic datasets. Therefore, we investigate the dependency of the DEV estimation results from intrinsic parameters like the AR order $p$ and window size $n_{\rm w}$.\\
Furthermore, we discuss the crucial role of proper detrending of the time series to get unbiased results via the DEV analysis. In particular, a proper detrending of the $CaCO_3$ concentration including the Eocene-Oligocene transition, leads to DEV results that promise new insights into the mathematical description of the transition: Complex conjugated eigenvalues approaching an absolute value of unity towards the Eocene-Oligocene transition $\sim\SI{34}{mya}$ imply the onset of periodic dynamics via a Neimark-Sacker bifurcation in a discrete model approach (i.e. a Hopf bifurcation in a continuous model ansatz)~\cite{a:kuznetsov,a:xin}. Taking into account that a lowering in atmospheric $CO_2$ acted as a trigger~\cite{a:DeConto2003} of the Eocene-Oligocene transition, the $CO_2$ concentration might play the role of a control parameter of the system. Considering this and keeping in mind the described Oligocene's heartbeat~\cite{a:Paelike2006}, the complex conjugated DEV results fit well into the qualitative picture of the Eocene-Oligocene climate transition. These results shed new light onto the underlying mechanisms: A Hopf bifurcation might be involved in the certainly more complex transition dynamics from the Eocene into the Oligocene climate state. Note that these insights are gained directly from empirical climate data and the results are robust under parameter variations (cf. Appendix \ref{subsec: appendix greenhouse result robustness}). In that sense, our findings might be a step forward to a better understanding of both, the Eocene-Oligocene transition and today's climate state. The results can help to improve existing climate models and favour those models which include a mathematical Hopf or Neimarck-Sacker bifurcation like e.g. the models of Plociniczak~\cite{a:Plociniczak2020} or Ghil and Tavantzis~\cite{a:ghil1984}. It may also inspire further discussion on the modelling of climate phenomena with nonlinear oscillators~\cite{a:Riechers2022}.\\
The remainder of the article is as follows: In section \ref{sec: theory} we summarize the linear stability formalism of iterated maps in subsection \ref{subsec: review theory} and introduce the simplified DEV estimation and a heuristic quality measure of the bifurcation classification in subsections \ref{subsec: method} and \ref{subsec: quality}, respectively. Afterwards we present our results in section \ref{sec: Results}. We describe three models in subsection \ref{subsec: datasets} which we use for methodological studies of the bifurcation classification quality and the DEV estimation precision in subsections \ref{subsec: quality results} and \ref{subsec: stationary DEV}, respectively. In subsection \ref{subsec: online DEV} we analyse the importance of proper detrending for the DEV estimation. A proof of principle for the stability estimation of limit cycles via the DEV approach and the alternative drift slope~\cite{a:Hessler2021} is presented in subsection \ref{subsec: results limit cycles}. In section \ref{sec: results climate} we employ the simplified DEV estimation to empirical $CaCO_3$ concentration data, which represent a climate proxy of the Eocene-Oligocene transition. Finally, we give a conclusion in section \ref{sec: conclusion}.\\

\section{\label{sec: theory}Theory}
In subsection \ref{subsec: review theory} the theory of the stability of iterated maps is briefly summarized as it is necessary for the numerical approach presented in subsection \ref{subsec: method}. The methodological investigation is completed by application of a specific quality measure for the classification of the bifurcation type that is introduced in subsection \ref{subsec: quality results}.
\subsection{\label{subsec: review theory} Review: Stability of iterated maps}
Without loss of generality we consider a two dimensional iterated map of the form
\begin{align}
\begin{aligned}
x_{\text{n+1}} &= f(x_{\text{n}}, y_{\text{n}}) \\
y_{\text{n+1}} &= g(x_{\text{n}}, y_{\text{n}}) 
\end{aligned}
\end{align}
with subsequent time series entries $x_n$, $x_{n+1}$ and discrete-time evolution functions $f$ and $g$. Linear stability of iterated maps is investigated by applying a small perturbation $\left\vert\eta^{\text{i}}_{\text{n}}\right\vert\ll 1$ with $i \in \lbrace x,y\rbrace$ to the fixed point $(x^*,y^*)$ and expand it into a Taylor series up to the first order. The result can be written as
\begin{align} \label{Eq: jacobian}
\vec{\eta}_{\text{n+1}} = \left( \begin{array}{c} \eta^{\text{x}}_{\text{n+1}} \\ \eta^{\text{y}}_{\text{n+1}} \end{array} \right) &= 
\left.\begin{pmatrix}
 \dfrac{\partial f}{\partial x} &  \dfrac{\partial f}{\partial y} \\
 \dfrac{\partial g}{\partial x} &  \dfrac{\partial g}{\partial y} 
\end{pmatrix}\right\vert_{(x^{\text{*}},y^{\text{*}})}
\cdot 
\left( \begin{array}{c} \eta^{\text{x}}_{\text{n}} \\ \eta^{\text{y}}_{\text{n}} \end{array} \right)\\ &=
\underline{\underline{J}}\vert_{(x^{\text{*}},y^{\text{*}})} \cdot \vec{\eta}_{\text{n}} .
\end{align}
The development of the perturbation in the next time step is defined by the eigenvalues $\lambda_{\text{i}}$ of the so-called Jacobian matrix $\underline{\underline{J}}$. This can be seen by diagonalization of the Jacobian $\underline{\underline{J}} = \underline{\underline{E}} \underline{\underline{\Lambda}}\underline{\underline{E}}^{-1}$. It follows
\begin{align}\label{Eq: scaled jacobian}
\vec{\eta}'_{n+1} =\underline{\underline{\Lambda}}\cdot \vec{\eta}'_n,
\end{align}
with the diagonal matrix $\underline{\underline{\Lambda}}$ of eigenvalues, the matrix $\underline{\underline{E}}$ of right eigenvectors and the scaled perturbations $\vec{\eta}' = \underline{\underline{E}}^{-1}\cdot \vec{\eta}$.~\cite{b:strogatz}\\
Eigenvalues with $|\lambda_{\text{i}}| < 1$ indicate a decay of perturbations, whereas perturbations grow for eigenvalues with $|{\lambda_{\text{i}}}| > 1$.\\
With permanent random noise $\eta^{\text{x}}_{\text{n}} = x_{\text{n}} - x^*$ and $\eta^{\text{y}}_{\text{n}} =y_{\text{n}} -y^*$ around a fixed point, equation \ref{Eq: jacobian} results in
\begin{align}
\vec{x}_{\text{n+1}} - \vec{x}^* &= J\vert_{(x^{\text{*}},y^{\text{*}})} \cdot (\vec{x}_{\text{n}} - \vec{x}^*) \\
\Leftrightarrow \vec{x}_{\text{n+1}} &= J\vert_{(x^{\text{*}},y^{\text{*}})} \cdot \vec{x}_{\text{n}}  \underbrace{-J\vert_{(x^{\text{*}},y^{\text{*}})} \vec{x}^*+ \vec{x}^*}_{\vec{x}_{\text{offset}}} \\
\Leftrightarrow \vec{x}_{\text{n+1}} &= J\vert_{(x^{\text{*}},y^{\text{*}})} \cdot \vec{x}_{\text{n}} +\vec{x}_{\text{offset}} .\label{Eq: numerical base}
\end{align}
\subsection{\label{subsec: method}Numerical procedure}
The connection between the high dimensional time evolution equation \ref{Eq: numerical base} and a one dimensional time series is established by Takens' theorem for stochastic systems presented in~\cite{a:stark}. The Takens delay embedding with dimension $d$ accounts for the multi-dimensionality of the system. The delay embedding results in the specific form
\begin{align}
 \vec{z}_{\text{n+1}} &= \left( \begin{array}{c} x_{\text{n+1}} \\ x_{\text{n}} \end{array} \right) \\
 &= \begin{pmatrix}
 j_1 & j_2 \\
 1 &  0 \end{pmatrix}
 \left( \begin{array}{c} x_{\text{n}} \\x_{\text{n-1}} \end{array} \right) + \left( \begin{array}{c} z_{\text{offset}} \\ 0 \end{array} \right) \label{Eq: jacobian_embed_coefficients}\\ 
 &= J_{\text{embed}} \cdot \vec{z}_{\text{n}} +\vec{z}_{\text{offset}}. \label{Eq: jacobian_embed}
\end{align}
for equation \ref{Eq: numerical base} with the delay embedding variables $\vec{z}_{\text{n}} = (x_{\text{n}},x_{\text{n-1}},...,x_{\text{n-d+1}})$ from a time series $X_t$. Takens' theorem states that the invariant features of the original system are conserved in the delay embedded space apart from a variable transform and thus, they are accessible due to this embedding procedure. In order to gain information about the eigenvalues of the Jacobian it is necessary to estimate the coefficients $j_{1,...,d}$ in the first row of the Jacobian. This estimation is done rather simple by application of an autoregression scheme
\begin{align}\label{Eq: general AR}
x_{n+1} = \sum_{i=0}^{p-1} j_{i+1}\cdot x_{n-i\tau} + z_{\text{offset}}
\end{align}
with the order $p \equiv d$. The autoregression assumes that subsequent time series values depend on past values weighted with the static coefficients $j_{1,...,d}$. Equation \ref{Eq: general AR} corresponds to the first row of equation \ref{Eq: jacobian_embed_coefficients} and the coefficients are numerically estimated via a least squares fit. An optimal embedding dimension $d$ can be found with the false nearest neighbour (fnn) algorithm described in detail in~\cite{b:kantz, a:kennel}. It is also summarized in Appendix \ref{subsec: fnn}.
The optimization of the time lag $\tau$ was not a crucial point in practise because of the AR($p$) based approach which relies on data past in time. Though, basically a proper time delay for an embedding could be estimated via the average distance from diagonal (add) algorithm~\cite{a:rosenstein} (cf. also Appendix \ref{subsec: add}). In contrast, an S-map estimation relies on proper unfolding of the underlying attractor~\cite{ip:deyle}.\\
The DEV is calculated for an ensemble of stationary datasets over a range of its specific control parameter. For each subsequent stationary set the control parameter is slightly increased up to its critical value. This allows to investigate the capacity of the DEV approach without the problem of a stationary window approximation that could create problems in an on-line implementation.\\
An on-line application needs rolling time windows~\cite{b:zivot,mt:ehebrecht}. The system state is assumed to be stationary for sufficiently small time windows and thus, the characteristic measures can be separately calculated in each rolling window. Possible problems are discussed with help of the online quality measures of subsection \ref{subsec: quality}.
\subsection{\label{subsec: quality}Prediction quality measures}
The quality of the numerical results depends on the right parameter choice. In our online application studies the most important parameters are a suitable embedding dimension $d\equiv p$ and the size of the rolling time windows $n_{\rm w}$. In practise the time lag was not an essential parameter for the presented approach, because the actual shadow attractor of the Takens embedding is not used in the numerical procedure. The quality measures $Q$ are designed especially for the analysis of the method's classification skill of bifurcations. A flip bifurcation of an iterated map is accompanied by a purely real dominant eigenvalue $\lambda$ approaching minus one~\cite{b:strogatz}. A Hopf bifurcation is indicated by two complex conjugated eigenvalues $|\lambda_{1,2}| = 1$~\cite{a:xin, a:kuznetsov} and the purely real dominant eigenvalue of other bifurcations, e.g. the fold bifurcation of one of the synthetic datasets in subsection \ref{subsec: datasets}, approaches one in the vicinity to the bifurcation point.
The quality measures $Q$ shall evaluate if the dominant eigenvalues approach the theoretically known threshold values of the simulated bifurcation types in the Gaussian plane. Furthermore, the difference of the dominant eigenvalues to the critical threshold should tend to decrease later in time in order to reconstruct a clear trend without random fluctuations in the Gaussian plane. A good classification skill should result in a rather small sum of difference values with a higher weight for difference values of eigenvalues that are calculated later and near to the bifurcation point. These methodological idea results in the sum of all absolute threshold - eigenvalue differences that are exponentially weighted in time relative to the absolute number $N$ of time series entries:
\begin{align}
Q_{\text{fold}} &= \frac{1}{N}\sum_{t=1}^N \vert 1 - \hat{\lambda}_t \vert \cdot \exp\left(\frac{\gamma \cdot t}{N}\right)\\
Q_{\text{flip}} &= \frac{1}{N}\sum_{t=1}^N \vert -1 - \hat{\lambda}_t \vert \cdot \exp\left(\frac{\gamma \cdot t}{N}\right) \\
Q_{\text{Hopf}} &= \frac{1}{N}\sum_{t=1}^N \vert \vert \hat{\lambda}_t \vert - 1 \vert \cdot \exp\left(\frac{\gamma \cdot t}{N}\right)  .
\end{align}
The hat denotes estimates. The weighting factor $\gamma$ is chosen in order to provide the best visualisation. The measure is not a real quantitative measure, but provides an idea of the classification quality with the highest classification skill for a small $Q$-factor and the lowest classification skill for the highest $Q$-factor. Also very small $Q$-factors are not desired, because all eigenvalues $\lambda \approx \lambda_{\text{crit}}$ and a possible trend is not well visualised.
\section{\label{sec: Results}Methodological studies}
The eigenvalue estimation approach based on the instructions in subsection \ref{subsec: method} is applied to three different models in this section. The models are presented in subsection \ref{subsec: datasets}, the bifurcation classification quality is discussed in subsection \ref{subsec: quality results} and the results of the DEV approach applied to stationary data subsets are shown in subsection \ref{subsec: stationary DEV}. In subsection \ref{subsec: online DEV} the crucial importance of proper detrending for on-line estimation purposes is demonstrated. Additionally, we discuss as a proof of principle how to use the DEV estimation and comparable measures to estimate the stability of limit cycles. The limit cycle model as well as the procedure are presented in subsection \ref{subsec: results limit cycles}.
\subsection{\label{subsec: datasets}Mathematical models}
Three datasets, each of which provides a different bifurcation type, are chosen analogous to~\cite{a:grziwotz}. The models are briefly introduced in the subsections \ref{subsubsec: ricker type}, \ref{subsubsec: henon} and \ref{subsubsec: hopf} and shown in figure \ref{fig: data} with control parameters that are changed linearly over time. In subsection \ref{subsec: stationary DEV} the datasets are simulated each individually for a fixed control parameter over the same parameter range as schematically suggested by the green stars in subfigure \ref{fig: data}. The initial data points are cut to avoid transient effects. This procedure ensures for stationary datasets.
\subsubsection{\label{subsubsec: ricker type}The Ricker-type model}
The discrete Ricker-type model, taken from~\cite{a:dakos17} and slightly modified, is developed to describe the population of a wide range of organisms like fish, birds and insects. The model is
\begin{align}\label{Eq: ricker}
x_{n+1} = x_n \exp\left( r-\varpi\cdot x_n\right) - F\frac{x_n^2}{x_n^2+\nu^2}+ \zeta\cdot \omega_{\text{noise}}
\end{align}
with a standard Gaussian white noise process $\omega_{\text{noise}}$, the initial condition $x_0 = 7.5$ and the parameters $\varpi = 0.1$, $\nu = 0.75$, $r=0.75$ and the noise amplitude $\zeta = 0.0375$.  In figure \ref{fig: data}(a) the control parameter $F$ is linearly increased in the range $[0,2]$ to visualize the changing dynamics of the model over $10^4$ time steps. Stationary versions of $2\cdot 10^4$ data points are generated with fixed values of $F$ in the same range for the basic analysis of the DEV approach. The critical threshold is at $F\approx 1.8$~\cite{a:dakos17}. The resulting time series is shown in blue in figure \ref{fig: data}(a) and the stationary versions that are used for the analysis are schematically signed by green stars.
\subsubsection{\label{subsubsec: henon} The Hénon model}
The two dimensional Hénon map time series shown in figure \ref{fig: data}(b) and written
\begin{align} \label{Eq: henon}
\begin{aligned}
x_{n+1} &= 1-a\cdot x_n^2 +y_n +\zeta\cdot \omega_{\text{noise}} \\
y_{n+1} &= b \cdot x_n
\end{aligned} 
\end{align}
with the parameter $b=0.3$, the control parameter $a$ and a standard Gaussian white noise process $\omega_{\text{noise}}$ undergoes a flip bifurcation at time $t\approx 9100$. The white noise process scales with $\zeta = 0.0025$ and the initial condition is $(x_0,\ y_0) = (0.3,\ 0.3)$. The control parameter $a$ is linearly increased from $0.1$ to $0.4$ over a range of $10^4$ time steps, shown in green in figure \ref{fig: data}(b). The control parameter's threshold value is $a = 0.3675$, depicted as green dashed horizontal line~\cite{a:aybar13}. The green stars in figure \ref{fig: data}(b) account for the stationary versions of the model with $2\cdot 10^4$ data points for a fixed control parameter $a$ in the range from $0.1$ to $0.6$.
\subsubsection{\label{subsubsec: hopf} The Hopf model}
\begin{figure*}
\centering
    \includegraphics[width =1.\textwidth]{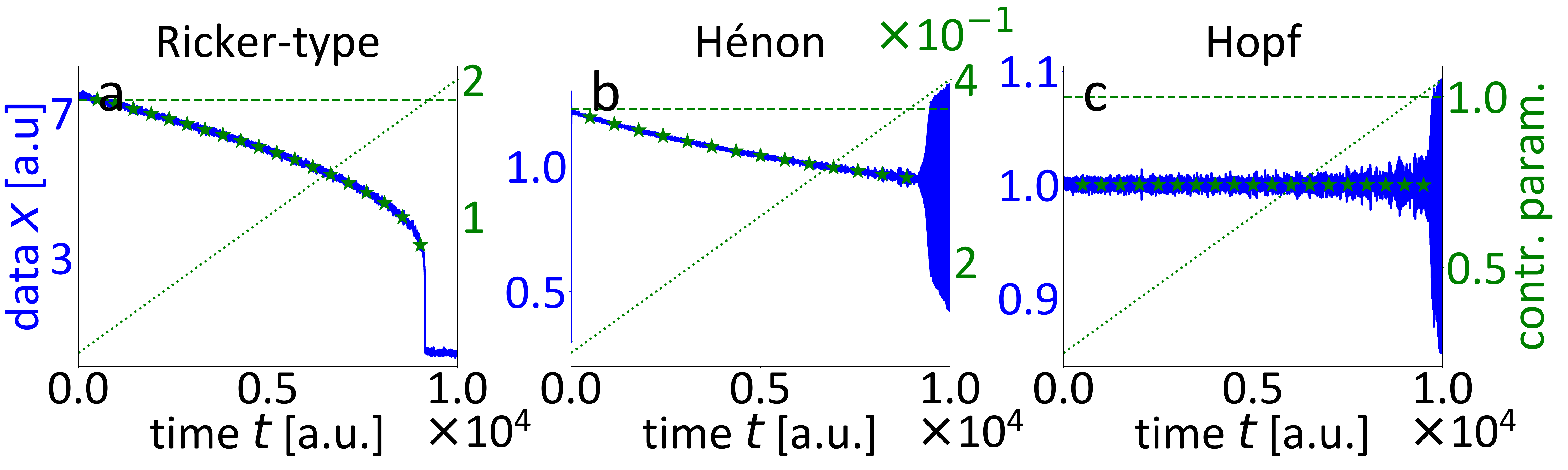}
    \caption[Time series of the three models.]{
(a) The time series of the Ricker-type model in equation \ref{Eq: ricker} over $10^4$ time steps. The parameters are listed in the running text. At approximately time $9100$ a fold bifurcation occurs. The corresponding threshold value of the control parameter  $F \approx 1.8$, drawn as green dashed horizontal line, is reached around this time as shown by the crossing green dotted line.\\
(b) The time series generated by the Hénon map in equation \ref{Eq: henon} is calculated over $10^4$ time steps. For parameter details, see the running text. At around $9100$ time steps a period doubling bifurcation occurs. The corresponding threshold value of the control parameter  $a = 0.3675$, drawn as green dashed horizontal line, is reached around this time, shown by the crossing green dotted line. The large initial deviations are caused by an initial transient of the simulation.\\
(c) The time series generated by the Hopf model in equation \ref{Eq: hopf} over $10^4$ time steps is shown in blue. At around $t=9300$ a Hopf bifurcation occurs. The corresponding threshold value of the control parameter  $u = 1$, drawn as green dashed horizontal line, is reached around this time, shown by the crossing green dotted line. More parameter details can be found in the running text.\\ 
(a-c) The green stars visualize the creation of stationary datasets for the analysis in subsection \ref{subsec: stationary DEV}. The analysis is done for stationary time series of about $2\cdot 10^4$ data points. Each stationary time series corresponds to an expanded version of the green star marked data. The control parameter is increased linearly between subsequent simulations.}
\label{fig: data}
\end{figure*}

The discrete two dimensional model with a Hopf bifurcation is taken from~\cite{b:kantz} and slightly modified. The time series in figure \ref{fig: data}(c) is calculated for the equations
\begin{align}\label{Eq: hopf}
\begin{aligned}
x_{n+1} &= \frac{1}{\sqrt{2}}\left(\vphantom{(x_n-1)^2 - 2(y_n -1 )^2} u (x_n -1 ) + (y_n - 1) \right.\\ &\left. +(x_n-1)^2 - 2(y_n -1 )^2\right) +1 +\zeta\cdot \omega_{\rm noise}\\
y_{n+1} &= \frac{1}{\sqrt{2}} \left(\vphantom{(x_n-1)^3} - (x_n-1) + u (y_n-1) \right. \\ &\left. +(x_n-1)^2 - (x_n-1)^3\right) +1 -2(y_n-1)^2 
\end{aligned}
\end{align}
with a standard Gaussian white noise process $\omega_{\rm noise}$, the parameter $\zeta = 0.0025$ and the initial condition $(1,\ 1)$. The  control parameter $u$ is linearly increased in the interval $[0.25, 1.05]$ over $10^4$ time steps. The resulting dataset is shown in figure \ref{fig: data}(c). The green stars indicate schematically the calculation of stationary time series with $2\cdot 10^4$ data points for fixed control parameter $u$ for the basic analysis of the DEV approach in subsection \ref{subsec: stationary DEV}.
\subsection{\label{subsec: quality results} Bifurcation classification quality}
\begin{figure*}
    \includegraphics[width = \textwidth]{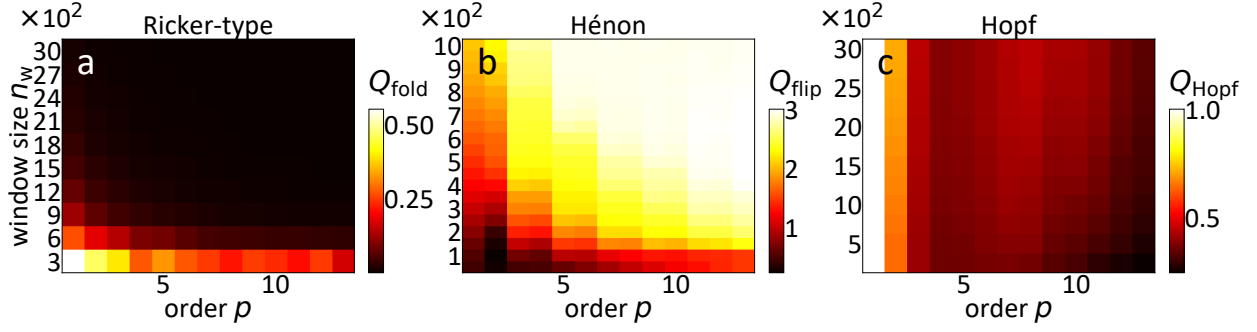}
    \caption[Quality tests of bifurcation type classification for the Ricker-type, the Hénon and the Hopf model]{The quality measures $Q$ for the bifurcation type classification are evaluated for the three investigated models (cf. equations \ref{Eq: ricker}, \ref{Eq: henon} and \ref{Eq: hopf}) for a wide range of time window sizes $n_{\rm w}$ and AR($p$) orders $p$. The quality measures $Q$ are defined in subsection \ref{subsec: quality}. They are a measure of scattering effects in the complex plane which do not represent the theoretically expected classification trend. A smaller quality measure corresponds to a better bifurcation type classification. $Q_{\rm fold, flip}$ are calculated with a time weight factor $\gamma = 1$, the calculations for the Hopf model were performed with $\gamma = 5$ for better distinguishing of the colormap range. Further details of the interpretation can be found in the running text.}
\label{fig: quality measure}
\end{figure*}

A detailed parameter scan with the quality measures $Q$, introduced in subsection \ref{subsec: quality}, is performed in order to show possible problems of the DEV estimation in an on-line moving window approach. Such a technique relies on the quasi-stationarity of each time window: That means that the stochastic dynamical process underlying the data does not change at all in any parameter over one time window range. The results are visualized in figure \ref{fig: quality measure}. Increasing order~$p$ and increasing window size~$n_{\rm w}$ stabilize the DEV trend of the Ricker-type model in the complex plane. It is usually convenient to choose the time windows as small as possible in order to save data, account for stationarity in the window and to obtain a fast trend response of the calculated measures~\cite{b:zivot, mt:ehebrecht}. Besides, the optimal order for relatively small stable time windows corresponds to the dimension that is suggested by the fnn algorithm (cf. Appendix \ref{subsec: fnn}). For very small time windows $n_{\rm w}\lessapprox 300$, the estimation of the DEV is unstable. For big time windows the quality measure is $Q_{\rm fold}\lessapprox 0.1$. That means that all eigenvalues lie in the vicinity of the EV that characterizes the bifurcation and the trend towards this value is not well resolved anymore in the Gaussian plane. This is probably an artefact, because for big time windows the constraint of the stationarity is violated more and more severely. This explanation is supported by the observation, that the time series of the Ricker-type and the Hénon model exhibit decreasing mean and the quality measure for the Hénon model is also affected by big time windows: For bigger time windows more and more EV are estimated in the vicinity of one instead of minus one and the trend performance is worse according to the quality measure. The quality measure for the Hopf model in figure \ref{fig: data}(c) is not sensitive to the choice of the time window, but to the AR order $p$. The Hopf time series does not exhibit a drift in the mean. This observation further supports the hypothesis that the drift in the mean causes artefacts for big time windows.\\ 
The false next neighbour algorithm tends to $d = 3$ for the Hénon model (cf. Appendix \ref{subsec: fnn}). The quality measure also seems to provide good classification results in a small range of time window sizes $n_{\rm w}\approx 150$ data points. Even better results in a wider time window range are suggested by $Q_{\rm flip}$ for $p=2$ which corresponds to the original dimensionality of the model.\\ 
The best results for the Hopf model are obtained for $p=4$. This is also the suggested embedding dimension of the fnn algorithm (cf. Appendix \ref{subsec: fnn}). The rather good quality measures $Q_{\rm Hopf}$ in the lower right corner for $p \geq 9$ are artefacts of unstable DEV scattering around the unit circle. However, the false next neighbour algorithm is a reliable tool to find a suitable order $p$ for the DEV analysis. The chosen optimal autoregression order $p$ for the further analysis is therefore chosen as 
\begin{enumerate}
    \item $p_{\rm Ricker}=d_{\rm Ricker}=3$,
    \item $p_{\rm Henon}=d_{\rm Henon} = 3$,
    \item $p_{\rm Hopf}=d_{\rm Hopf}=4$.
\end{enumerate}
as supposed by the false next neighbour algorithm to ensure consistency of the procedure.\\

\subsection{DEV estimation on stationary time series}
\label{subsec: stationary DEV}
\begin{figure*}[tp!]
    \includegraphics[width = 0.99\textwidth]{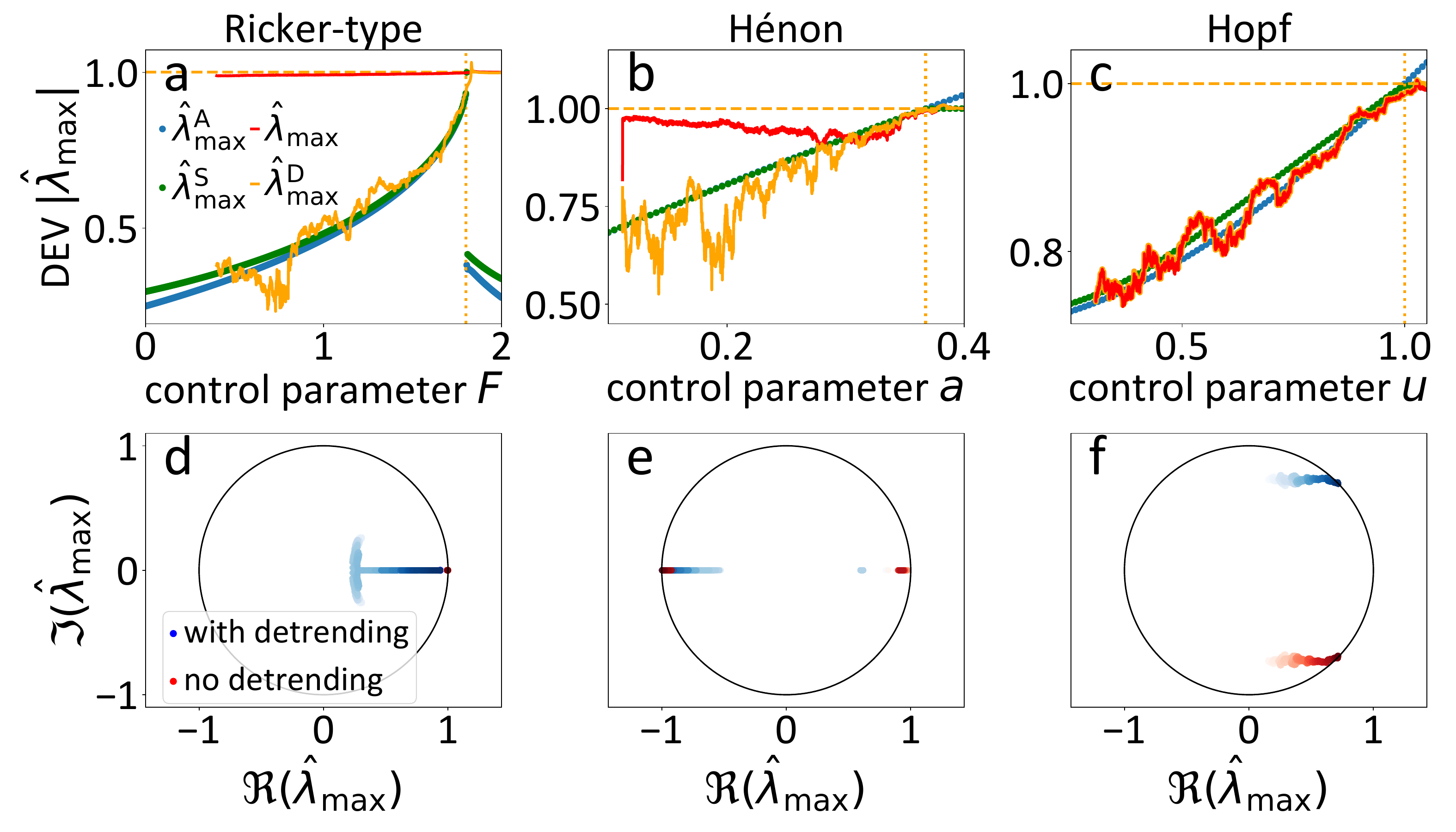}
    \caption[No DEV distortion without detrending]{(a-c) A detailed comparison of the DEV moduli estimates to the ground truth. $\hat{\lambda}^{\rm D}_{\rm max}$ and $\hat{\lambda}_{\rm max}$ denote estimates with and without preceding detrending of the data, respectively. The analytically calculated DEV and the DEV of the corresponding stationary data simulations  are given by $\hat{\lambda}^{\rm A}_{\rm max}$ and $\hat{\lambda}^{\rm S}_{\rm max}$, respectively. The results are shown for the Ricker-type model (a), the Hénon model (b) and the Hopf model (c). The parameters for the estimation processes are: Ricker-type model: $d = 3$, window size $n_{\rm w}$ ${\rm ws} = 2000$; Hénon model: $d = 3$, $n_{\rm w} = 400$; Hopf model: $d = 4$, $\rm ws = 700$. The DEVs $\hat{\lambda}^{\rm D}_{\rm max}$ with preceding detrending agree well with the expected analytical and stationary estimated eigenvalues. Without detrending the shifting mean in the Ricker-type and the Hénon model lead to strong distortions from the ground truth.\\
(d-f) Similar to the DEV moduli the DEVs in the Gaussian plane show distortions for the Ricker-type model (d) and the Hénon model (e). The time evolution is resolved from transparent to opaque points. (d) Almost all DEVs without detrending (red) accumulate at $\hat{\lambda}_{\rm max} \approx 1$. The detrended DEV estimates (blue) show instabilities for estimates early in time which can be avoided for bigger window sizes/smaller AR order $p$. (e) The Hénon model shows the most significant distortions in the Gaussian plane. The DEV estimates without detrending jump between $\hat{\lambda}_{\rm max} \approx -1$ and $\hat{\lambda}_{\rm max} \approx 1$ regardless of the considered time period. This makes it impossible to classify the bifurcation type. A detrending solves the issue: Only a few estimates early in time seem to be estimated unstable (cf. $\hat{\lambda}_{\rm max} \approx 0.6$). The DEV estimates of the Hopf model (f) are not biased. For illustration purposes we show one branch of complex conjugates of the raw/detrended data results.}
\label{fig: synthetics results}
\begin{minipage}{0.45\textwidth}
	\includegraphics[width = \textwidth]{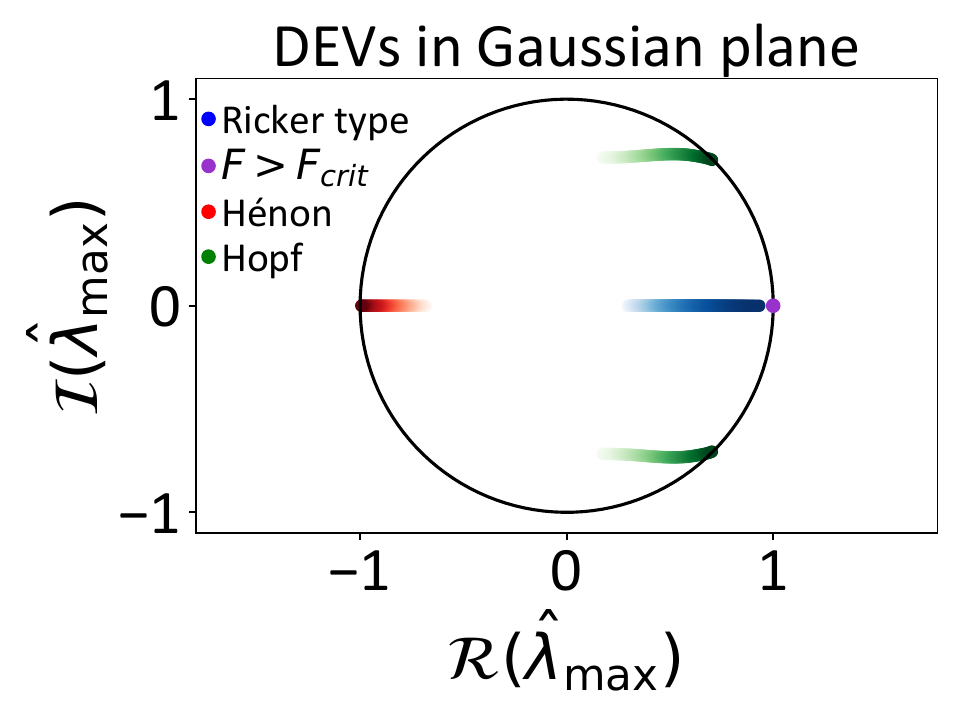}
\end{minipage}
\begin{minipage}{0.55\textwidth}
	\caption{The evolution of the DEV in the Gaussian plane is visualized with transparent to opaque points that correspond to datasets simulated with control parameters from the lower limit of the range to the upper limit, respectively. The DEVs are plotted until the critical control parameter is reached. The three different bifurcation types can be classified with the DEV trend in the Gaussian plane: The fold bifurcation of the Ricker-type model is indicated by the real DEV approaching one. The Hopf bifurcation is identified by the two complex conjugated DEVs approaching an absolute value of one and the flip bifurcation of the Hénon map is characterized through the DEV that approaches minus one. It should be mentioned that the trend of the Ricker-type DEV fits also for other bifurcations and is not sufficient to identify precisely a fold bifurcation. The darkorchid point is plotted apart from the Ricker-type DEV set, because it is the first DEV with $F\approx 1.805$ slightly behind the critical control parameter $F_{\rm crit}=1.8$.}
	\label{fig: AR complex plane trend}
\end{minipage}
\end{figure*}

The DEV estimation is performed on stationary sets of each model. Each dataset consists of $2\cdot 10^4$ data points to avoid possible artefacts caused by the window size $n_{\rm w}$ and non-stationary windows. The last $10^4$ points are used for the analysis to avoid errors due to an initial transient in the simulations. For the Ricker-type model $400$ datasets with fixed equidistantly distributed control parameters $F$ are generated in the range from $0$ to $2$. For the Hénon model $100$ datasets with fixed equidistantly chosen control parameters $a$ are calculated in the range $[0.1,0.6]$ and for the Hopf model $80$ datasets for fixed equidistantly distributed control parameters $u$ in the range $[0.25,1.05]$ are simulated. The results of the DEV analysis are shown in figure \ref{fig: synthetics results}. 
The estimated DEVs $\hat{\lambda}^{\rm S}_{\rm max}$ for the corresponding stationary datasets are shown as green dots. Moreover, the DEV estimates $\hat{\lambda}^{\rm S}_{\rm max}$ are compared to their analytical pendants. They are denoted by $\hat{\lambda}^{\rm A}_{\rm max}$. In case of the Ricker-type model the analytical fixed point is approximated by the mean values of stationary Ricker-type simulations for a discrete range of control parameters $F$ (cf. the procedure in~\cite{a:Hessler2022}). For comparison of the estimation precision these analytical DEVs are shown as blue dots in figure \ref{fig: synthetics results}. As theoretically expected it holds $\hat{\lambda}^{\rm S}_{\rm max} \approx \lambda^{\rm A}_{\rm max}$ for all models. All investigated models show a positive trend when the control parameters increase towards a critical value which is marked by the vertical dotted orange lines. In vicinity of that critical value the DEV approaches the stability threshold $|\lambda |= 1$ as indicated by the horizontal dashed orange line. The Hénon model saturates even before the critical value $a_{\rm crit}$ is reached, the Ricker-type slightly behind the critical parameter value $F_{\rm crit}$ and the Hopf model fits rather exact with the critical parameter value of $u_{\rm crit}$. Similar to the autocorrelation at a lag of one the DEV has an upper limit of one that provides a rough idea of the distance to a critical transition.\\
The DEV trend in the complex plane is presented in figure \ref{fig: AR complex plane trend}. The three different bifurcation types can be identified by their typical DEV time evolution that is resolved by a colour shading from transparent to opaque points. The fold and flip bifurcation of the Ricker-type and the Hénon model are identified by the evolution towards one and minus one, respectively, and the Hopf bifurcation is identified by complex conjugated eigenvalues that approach an absolute value of one. The DEVs are plotted up to the critical control parameter of each model. For the Ricker-type model the first DEV with $F \approx 1.805$ slightly behind the critical control parameter $F_{\rm crit} = 1.8$ is separately shown, because it demonstrates that the DEV trend in the Gaussian plane reaches the stability thresholds in the vicinity of the critical parameter values, but does not necessarily reach exactly unity prior to the critical threshold depending on the sampling rate and possibly disturbance e.g. by strong noise.
\subsection{\label{subsec: online DEV}On-line DEV estimation}
Prior to the on-line DEV AR($p$) estimation the variance $\sigma^2$ in each window is tested against stationarity to keep the analysis consistent. To this goal the variance in each window is estimated. In a next step the variance of moving subwindows of $n_{\rm sub} = 20$ is computed. The variance in a window is defined as stationary if no subwindow variance deviates from the window variance by a predefined percentile $\alpha$ prior to the bifurcation. The variance of the Hénon and the Hopf model do not deviate by more than \SI{1}{\percent} prior to the bifurcation point. The Ricker-type model deviates less than \SI{5}{\percent}. Admittedly, the test is a very simple and naive approach that uses a subjective condition $\alpha$ to classify stationary variance in a window, but it is suitable for the synthetic analysis performed here.\\
The DEV $\lambda_{\rm max}$ is estimated with and without linear detrending on each time window to illustrate the sensitivity of the estimation process to non-stationary data. 
The chosen parameters for the numerical estimation approach are given by\\
\begin{enumerate}
		\item Ricker-type model: $d = 3$, window size $n_{\rm w} = 2000$,
		\item Hénon model: $d = 3$, $n_{\rm w} = 400$,
		\item Hopf model: $d = 4$, $n_{\rm w} = 700$.
\end{enumerate}
The results are shown in figure \ref{fig: synthetics results}. The DEV $\hat{\lambda}^{\rm D}_{\rm max}$ that are estimated after detrending fluctuate around the analytical values as expected. Nevertheless, the absence of the detrending procedure leads to rather strong distortions of the absolute values of the DEV $\hat{\lambda}_{\rm max}$ for the Ricker-type and the Hénon model in figure \ref{fig: synthetics results}(a) and \ref{fig: synthetics results}(b), but not for the Hopf model in figure \ref{fig: synthetics results}(c). This is caused by the non-stationary mean of the former two, whereas the latter model does not exhibit a mean shift. Similar to the DEV modulus results, the DEVs in the Gaussian plane without prior detrending are distorted for the Ricker-type model, shown in figure \ref{fig: synthetics results}(d), and for the Hénon model, shown in figure \ref{fig: synthetics results}(e). The DEV estimates of the Ricker-type model without detrending do not show a well-resolved trend with $\hat{\lambda}_{\rm max}\approx 1$. The detrended version show only instabilities for early estimates. These instabilities can be typically circumvented by using bigger time windows and/or smaller AR order $p$.\\
Note that the distortions of the DEV estimates of the raw Hénon data have the most severe consequences. They jump between $\hat{\lambda}_{\rm max} \approx -1$ and $\hat{\lambda}_{\rm max} \approx 1$ regardless of the considered time period. Therefore, it is impossible to classify the flip bifurcation correctly. A detrending enables us to classify the flip bifurcation. Only some early DEVs are estimated as $\hat{\lambda}\approx 0.6$. Theses distortions might be due to small data windows or the chosen order $p$. For similar reasons as discussed for the DEV modulus, the Hopf model estimates in figure \ref{fig: synthetics results}(f) are not affected by improper detrending.

\subsection{Stability analysis of limit cycles}
\label{subsec: results limit cycles}
\begin{figure*}
    \centering
    \includegraphics[width=0.95\textwidth]{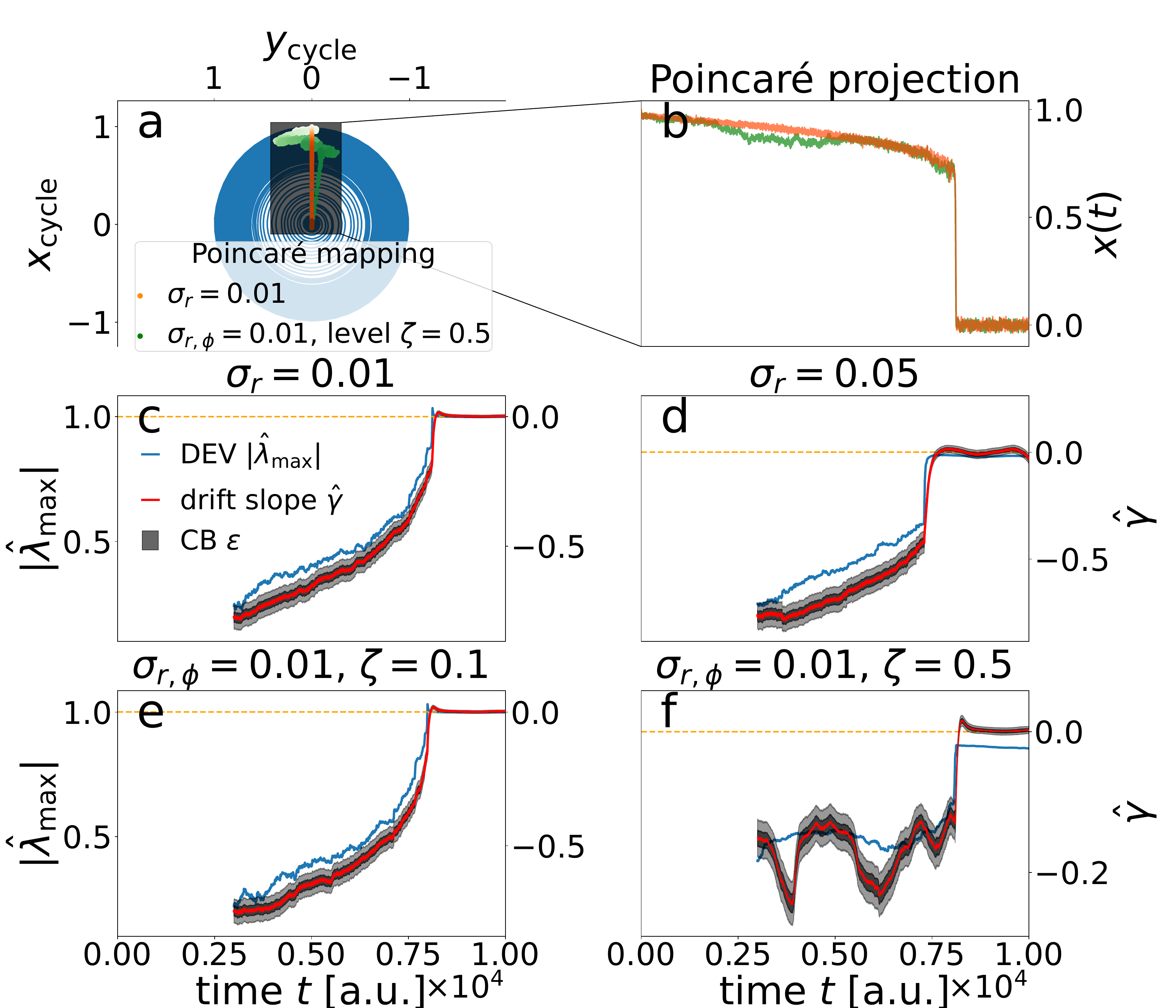}
    \caption{Data generation and analysis of limit cycles' stability. (a) The Cartesian state space with an example solution of equation \ref{eq: limit cycle}. The trajectory follows the attracting limit cycle with radius $\tilde{r}_0=1$ that shrinks continuously due to a linear control parameter shift of $\mu$ from $-0.05$ to $-0.3$, creating a spiral trajectory. Finally, it disappears in a radial saddle node bifurcation at $\mu_{\rm crit} =-0.25$. The trajectory collapses into the stable fixed point $(0,0)$. As an example the (approximated) Poincaré maps of the trajectory with $\sigma_r = 0.01$ and with $\sigma_{r,\phi}=0.01$ and $\zeta=0.5$ are shown. The Poincaré map approximation is based on a known period length of $50$ time steps. Using this information every 50th data point is sampled to construct the approximated Poincaré map. The time evolution is color-coded from white to orange/green. (b) The resulting time series of the Poincaré maps' projections onto the $x$-axis. (c-d) Stability analyses' results of the DEV $\hat{\lambda}_{\rm max}$ and the drift slope measure $\hat{\gamma}$ introduced in~\cite{a:Hessler2021,a:Hessler2022} for four different noise combinations and strengths in equation \ref{eq: limit cycle}. (c-e) Both measures exhibit clear positive trends towards their destabilization thresholds (i.e. one and zero, respectively) for small and strong noise in the radial component only or combined with weak disturbances of the angular velocity. In general, the drift slope and DEV estimates exhibit very similar behaviour.
    (f) Both stability measures are not able to resolve a destabilization trend for stronger disturbance of the angular velocity. Further analysis details can be found in Appendix \ref{subsec: appendix limit cycles}.}
    \label{fig: limit cycles}
\end{figure*}
In this subsection we evaluate the stability of limit cycles as a numerical proof of principle. We compare the DEV $\hat{\lambda}^D_{\rm max}$ to alternative stability estimates and discuss the crucial role of noise contributions to the angular velocity, i.e. the period length, respectively. Therefore, we generate realisations of the process~\cite{b:strogatz} 
\begin{align}
\label{eq: limit cycle}
    \odv{r}{t} &= \mu \tilde{r} +\tilde{r}^3 - \tilde{r}^5 + \sigma_r \cdot \text{d}W \\
    \odv{\phi}{t} &= 2\pi + \zeta \cdot \sigma_\phi \cdot \text{d}W
\end{align}
in polar coordinates $(\tilde{r},\phi)$ with a Standard-Wiener process $\text{d}W$, noise levels $\sigma_{r,\phi}$ and an extra noise parameter $\zeta\in\mathbb{R}^{\geq 0}$ to turn on and off noise contributions to the polar angle. The limit cycle destabilizes due to a fold bifurcation in the radial component when crossing the critical control parameter threshold $\mu_{\rm crit} = -\frac{1}{4}$. The destabilization in the simulations is triggered by tuning $\mu$ linearly from $-0.05$ to $-0.3$ and integrating $5\cdot 10^5$ steps over the time range $[0,10000]$ via the Euler-Maruyama method. The initial coordinates are $(\tilde{r}_0 = 1, \phi_0 = 0)$. We discuss four representative parameter combinations, namely
\begin{enumerate}
    \item $\sigma_r = 0.01$,
    \item $\sigma_r = 0.05$,
    \item $\sigma_r = 0.01$, $\zeta = 0.1$,
    \item $\sigma_r = 0.01$, $\zeta = 0.5$.
\end{enumerate}
The former two only include noise in the radial coordinate whether the latter two additionally incorporate noise in the polar angle. The generated $(\tilde{r},\phi)$-realisations are transformed in Cartesian coordinates $(x_{\rm cycle}, y_{\rm cycle})$. A Poincaré map is constructed in two steps:
\begin{enumerate}
    \item Every 50th data point $(x_{\rm PC},y_{\rm PC})$ is sampled which corresponds to one period $T$ of the process.
    \item The sampled points are projected onto the $x$-axis (note that for our $r_0,\phi_0$ it holds $y_{\rm PC}\equiv 0$ for $\zeta = 0$ and $y_{\rm PC}\approx 0$ for $\zeta \gtrsim 0$).
\end{enumerate}
In empirical studies the approximate period of the signal has to be known to derive a Poincaré map of the signal. This procedure is illustrated in figure \ref{fig: limit cycles}(a,b). In figure \ref{fig: limit cycles}(a) the blue process trajectory is attracted to a limit cycle with initial radius one which is continuously decreasing. This creates a spiral trajectory, before the limit cycle disappears in a radial saddle node bifurcation by which the system state falls down to the arising stable fixed point $(0,0)$. The Poincaré map coordinates of the first (orange dots) and fourth (green dots) parameter combinations are also shown in state space as an example. The time evolution is encoded from white to dark orange/green dots. These points are projected onto the $x$-axis which leads to the finally analysed time series presented in figure \ref{fig: limit cycles}(b). The Poincaré maps that are derived this way capture the information about the saddle node bifurcation of the radial component of the process in polar coordinates. Note that principally this holds w.l.o.g. for projections from hyperplanes with $y_{\rm PC}\neq 0$ as long as the fold bifurcation can be adequately resolved, which does not hold for
\begin{align}
(x_{\rm PC},y_{\rm PC})\xrightarrow[\phi\cdot T \to \frac{\pi}{2}]{} (0,1).
\end{align}
However, that has no consequences in practise, since the periodic process would be identified in $y_{\rm PC}$ and thus, projected onto the $y$-axis in this case.\\
The corresponding results of the DEV $|\hat{\lambda}_{\rm max}|$ are shown as blue lines in the figures \ref{fig: limit cycles}(c-f). Further details of the analysis, parameters and pre-processing can be found in Appendix \ref{subsec: appendix limit cycles}. The DEV $|\hat{\lambda}_{\rm max}|$ reaches its critical value one almost without discontinuity near the destabilization time $t_{\rm crit}\approx 8000$ 
\begin{enumerate}
    \item as long as intrinsic noise is only present in the radial component of the analysed process  (cf. figure \ref{fig: limit cycles}(c)),
    \item as long as the disturbances of the angular velocity/the period length are small, i.e. $\sigma =0.01$, $\zeta = 0.1$ in our examples (cf. figure \ref{fig: limit cycles}(e)).
\end{enumerate}
Stronger stochastic contributions in the radial component only lead to a destabilization trend with a discontinuous jump around the transition time as visible in figure \ref{fig: limit cycles}(d). Our results in figure \ref{fig: limit cycles}(f) suggest that the applicability of the numerical approach is mostly limited in the case of unknown disturbances in the angular velocity, since already noise levels $\zeta\cdot\sigma=0.005$ which are small compared to numerically stable ones in the radial component (at least a factor of ten for $\sigma_r = 0.05$) result in DEV $|\hat{\lambda}_{\rm max}|$ estimates without any destabilization trend. This can be easily understood by considering the time evolution of the Poincaré mapping coordinates from white to green dots of the fourth study case ($\sigma_r = 0.01$, $\zeta = 0.5$) in figure \ref{fig: limit cycles}(a). If the stochastic variations in angular velocity are unknown in an empirical study the scientist has to sample the Poincaré map with a time step that corresponds only to an approximated fixed period of the process. In our example this already leads to high deviations from the actual fixed hyperplane of the Poincaré map, i.e. the line $(x,0)$ with $x\in\mathbb{R}$ (which corresponds almost to the orange dots in figure \ref{fig: limit cycles}(a)), compared to the green dots. A practical solution of the problem can be achieved if the experimenter is able to measure the disturbances of the angular velocity of the process almost instantaneously. In such a situation the experimenter can either adapt the position of the assumed hyperplane accordingly imitating fixed angular velocity or correct the adjacent sampling step to be in line with a fixed hyperplane. This fixed plane might be defined e.g. by the previously sampled data (in our case this could be the previously mentioned line $(x,0)$ with $x\in\mathbb{R}$). In other words, basically a flexible sampling step depending on the observed disturbances in angular velocity has to be used. Nevertheless, note that, e.g. in our simulations the second approach of a fixed hyperplane leads to a bias in the radial dynamics with linear changes in the control parameter $\mu$. This is not the case for the first approach that guarantees a fixed period and thus, equidistant sampling of the changing radial dynamics.
Following one of the sketched practical procedures, the disturbance would enter significantly less or not at all into the subsequent numerical analysis of the Poincaré map anymore. However, situations with precisely measurable disturbance in the process period might be rare. Therfore, a third solution might be a statistical approach involving an ensemble of potential Poincaré hyperplanes.\\
Additionally, we calculate the drift slope $\hat{\gamma}$ using the Python package \textit{antiCPy}\cite{url:GitHessler2021, url:DocsHessler2021} which is introduced in~\cite{a:Hessler2021,a:Hessler2022} as an alternative stability measure shown by the red line with dark grey $\SIrange{16}{84}{\percent}$ and $\SIrange{1}{99}{\percent}$ credibility bands (CBs) to provide a comparison of the two approaches. Further analysis details are summarized in Appendix \ref{subsec: appendix limit cycles}. The destabilization threshold of the drift slope $\gamma$ is $\gamma_{\rm crit} = 0$. Both measures show in general very similar behaviour. 
\section{Anticipating Oligocene's climate heartbeat}
\label{sec: results climate}
\begin{figure*}
    \includegraphics[width = \textwidth]{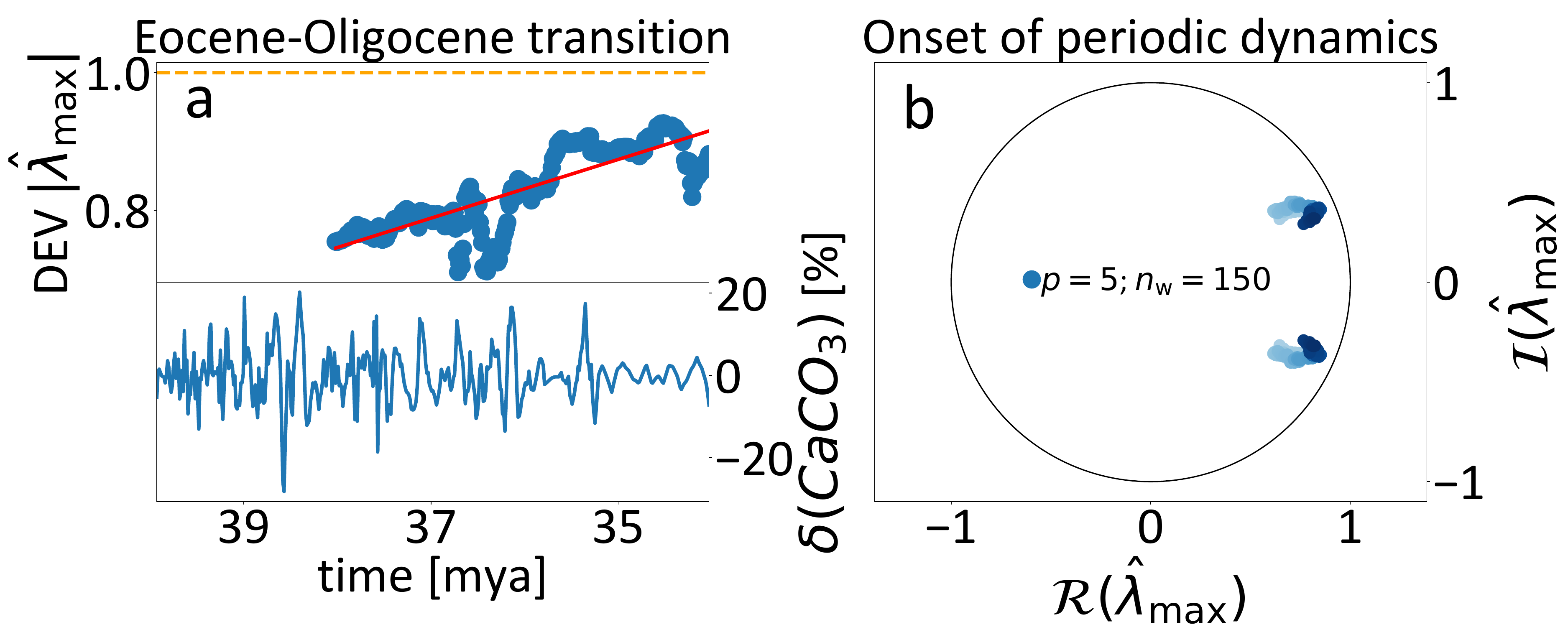}
\caption[]{(a) The detrended $CaCO_3$ is shown in the lower graph. The upper graph shows a positive trend in the DEV evolution over time, supported by the red linear fitted line. The line can guide the eye, but clearly cannot be interpreted as the real eigenvalue evolution law. The positive trend could give a hint to the upcoming transition at time $t \gtrsim \SI{3.4}{\cdot 10^7 mya}$.\\
(b) The DEV evolution in the Gaussian plane is shown. The time evolution is encoded from transparent to opaque dots. The occurrence of complex conjugated eigenvalue pairs approaching a modulus of one implies that a Neimark-Sacker bifurcation of iterated maps (i.e. a Hopf bifurcation in continuous time models) is involved in the Eocene-Oligocene transition. The findings agree well with the observed onset of periodic dynamics in the beginning of the Oligocene, even if the critical modulus of one is not exactly reached in the considered time interval. This might be due to data quality and/or the chosen time period. The last DEV estimates tend to decrease slightly in modulus. However, this is probably caused by the method parameter choice and the empiric data quality. The presented result is reproducible over a wide range of method parameters. Certain parameter combinations do not even show the slight decrease in modulus (cf. Appendix \ref{subsec: appendix greenhouse result robustness}).}
\label{fig: green house}
\end{figure*}

In paleoclimatology the $CaCO_3$ concentration extracted from foraminifera shell material in sea sediment cores is used as a proxy of sea water temperatures and the carbon cycle in ancient climate states~\cite{b:Gornitz}. We analyse such data from a pacific sediment core at spot DSDP1218 from Tripati et al.~\cite{d:Tripati2005, a:Tripati2005}. The data record covers the transition from a high temperature greenhouse climate, present in the Eocene, to the cooler icehouse state with the first formation of the Antarctic ice shield in the beginning of the Oligocene roughly $\SI{34}{mya}$. The $CaCO_3$ concentration is analysed over a range from $\SIrange{34.03}{39.93}{mya}$. We have to apply two preprocessing steps in close analogy to Dakos et al.~\cite{a:dakos08} (cf. Appendix \ref{sec: appendix greenhouse}). First, we interpolate the dataset, but keep the original total number of data points, since the AR($p$) estimation holds only for equidistantly sampled data. Second, since the non-stationarity of the time series is rather complex, we replace the previously used simple linear detrending method by subtracting a slow trend which is estimated via a Gaussian kernel filter. More details on the interpolation and the detrending scheme can be found in Appendix \ref{subsec: appendix data preparation} and \ref{subsec: appendix gauss detrend}, respectively. The detrended dataset is shown in the lower graph of \ref{fig: green house}(a). The critical transition occurs around time $t \gtrsim 34$ \si{mya}. The optimal order $p$ of the autoregression process is chosen to be equal to the optimal embedding dimension $d_{\rm opt}$ of the fnn algorithm~\cite{a:kennel} for the detrended data. The DEV estimation is performed for the parameters 
\begin{enumerate}
	\item order $p = d_{\rm opt} = 5$ (cf. fnn results in Appendix \ref{subsec: appendix fnn greenhouse}),
	\item $n_{\rm w} = 150$ (analogue to Grziwotz et al. \cite{a:grziwotz}).
\end{enumerate}
The absolute DEV trend is shown in the upper graph in figure \ref{fig: green house}(a). Of course, the real DEV evolution is not necessarily linear and typically unknown, but the positive DEV trend is illustrated by the linear fitted red line that shall guide the eye. The positive trend is in accordance with the results of Dakos et al.~\cite{a:dakos08} for the positive trend in the autocorrelation. Nevertheless, it should be mentioned that the standard deviation exhibits a negative trend (cf. Appendix \ref{subsec: appendix std greenhouse}), whereas evidence for critical slowing down is only provided for simultaneously positive trends in both, autocorrelation and standard deviation~\cite{a:Ditlevsen2010, a:scheffer12, a:dakos12}. Therefore, the provided DEV trend might be a more reliable indicator of an approaching bifurcation involved in the Eocene-Oligocene transition.\\
The time evolution of the DEV in the Gaussian plane is shown in figure \ref{fig: green house}(b) with time encoded from transparent to opaque dots. The computed DEVs are complex conjugated with
\begin{align}
|\lambda^{(1,2)}_{\rm max}|\xrightarrow{t\to \SI{34}{mya}}{} 1
\end{align}
which implies the onset of periodic dynamics during the transition from late Eocene to early Oligocene due to a Neimark-Sacker bifurcation~\cite{a:kuznetsov, a:xin} in the framework of iterated maps (i.e. equivalent to a Hopf bifurcation~\cite{b:strogatz} in a continuous mathematical description). These findings are accompanied by visual inspection of approximate unfoldings of shadow manifolds in three dimensions in Appendix \ref{subsec: shadow manifold}. We have to keep in mind that an unfolding in three dimensions is below the optimal embedding dimension $d_{\rm opt}=5$. Thus, the results should be judged carefully. Nevertheless, some qualitative observations seem to agree rather good with the DEV estimation results: Perturbations seem to be damped in spiral-like dynamics with a drift of a bound state prior to the onset of the Oligocene. Additionally, we consider embeddings of the whole time series. They might resemble transient drift dynamics  roughly $\SI{40}{mya}$ and after the Eocene-Oligocene transition around $\SI{34}{mya}$. We extensively tested the robustness of our results against variation of the methods' intrinsic parameters and detrending bandwidths $\tilde{\sigma}$ in Appendix \ref{subsec: appendix greenhouse result robustness}. The results are reproducible roughly between $d = \SIrange[]{3}{7}{}$ for window sizes $n_{\rm w} = \SIrange[]{100}{200}{}$. The detrending turns out to be the most crucial dependency. The results are roughly reproducible for all bandwidths, but scatter more in the Gaussian plane for a weaker detrending (i.e. bandwidth $\tilde{\sigma}=20$). This is an expected result, since the AR$(p)$ requires stationary data which becomes compromised by improper detrending of an underlying slow trend in the time series.\\

\section{\label{sec: conclusion}Conclusion}
The periodicity of glacial-interglacial cycles~\cite{ic:Maslin,a:Boers2022} is a well-known characteristic of the climate over the last $\SI{1}{Ma}$ of the Quaternary period~\cite{ic:Ehlers2011} which lasts from $\SI{2.6}{mya}$ to the present day. The Quaternary period as well as the Eocene and Oligocene are periods of the Cenozoic era~\cite{a:stratigraphy}. The Eocene was characterized by warm climate and the absence of ice in both polar regions. It was a so-called greenhouse climate state~\cite{a:Evans2018}. The beginning of the icehouse Oligocene climate is marked by the exceptional event of the first glaciation of Antarctica~\cite{a:Tripati2005, a:Galeotti2016}. More details of the climate conditions in the Oligocene are obtained through evaluation of a Pacific sediment core by Pälike et al.~\cite{a:Paelike2006}. The extracted data covers $\SI{13}{Ma}$ of the Oligocene, revealing periodic changes in the carbon cycle and glacial-interglacial cycles similar to today's Quaternary period. Paelike et al.~\cite{a:Paelike2006} call this pronounced periodicity ``The heartbeat of the Oligocene climate system''. The exact mechanisms of that heartbeat are not fully understood, but the authors show that a Milankovitch-type~\cite{b:milankovitch1920} astronomical forcing of the carbon cycle, the atmospheric $CO_2$ concentration, the temperature and glaciations play an important role. Furthermore, DeConto et al.~\cite{a:DeConto2003} observed that the actual Eocene-Oligocene transition is predominantly triggered by a significant lowering of atmospheric $CO_2$ concentration. In mathematical terms, both findings combined could imply a Hopf bifurcation from warm Eocene climate to periodic Oligocene climate conditions by a control parameter change, i.e. the lowering of the $CO_2$ concentration.
Our DEV analysis suggests exactly such type of transition and is directly based on empirical climate data. The results are robust under parameter variations (cf. Appendix \ref{subsec: appendix greenhouse result robustness}). Additionally, we considered approximate unfoldings of the data in three dimensions in Appendix \ref{subsec: shadow manifold}. We have to keep in mind that we could not choose the optimal embedding dimension for inspection by eye. Nevertheless, some spiral-like behaviour with a shifting bound state prior to the Eocene-Oligocene transition might fit well into the picture that is drawn by the DEV analysis. Furthermore, there seem to be hints to transient drifts in the beginning and the end of the whole dataset.\\
In Grziwotz et al.~\cite{a:grziwotz} the DEV analysis was performed without proper detrending of the empirical data. For this reason, the uprising periodic dynamics were hidden by the impact of the slow trend even if the study introduced the powerful DEV approach using S-maps while keeping the preprocessing simple. 
This underlines the outstanding importance of proper detrending for the DEV analysis. Previously, we demonstrated the crucial importance of proper detrending considering synthetic data in section \ref{subsec: online DEV}: When the DEV estimation is applied to time series via an on-line rolling window approach, missing detrending can lead to significant biased DEV estimates. The strongest effect is observed in the Gaussian plane, where estimates seem to scatter due to an unstable estimation procedure.\\
Further numerical investigations that are presented in this article confirm that the S-map based DEV estimation approach, that was firstly introduced by Grziwotz et al.~\cite{a:grziwotz}, can be simplified by an autoregressive estimation of the parameters $j_{1,2,...}$ of the Jacobian~$J_{\rm embed}$. In contrast to other studies, the analysis of the DEV as leading indicator is done with stationary datasets to avoid misleading artefacts due to non-stationarity. The results confirm that the basic concept of the numerical DEV estimation can provide useful information about the stability of the system and in particular the type of an uprising bifurcation. Similar to the autocorrelation it has an upper limit of one and thus, gives some rough intuition of the vicinity of a bifurcation. In comparison to the standard leading indicators of autocorrelation and standard deviation it has to be emphasized that the DEV approach is able to classify various bifurcation types like flip, Hopf and other bifurcations.\\
With a methodological quality measure that is introduced in this paper, it is shown that the false next neighbour algorithm provides an optimal embedding dimension that corresponds to a suitable autoregression order $p$ with some small discounts for the Hénon model.\\
Finally, as a proof of principle we discussed how the stability of limit cycles can be estimated via the DEV or the alternative drift slope measure~\cite{a:Hessler2021, a:Hessler2022,url:GitHessler2021, url:DocsHessler2021}. The period length of the process is crucial to know in order to construct Poincaré maps from empirical data. Our studies show that the stability estimates for a limit cycle are more robust against disturbances in the radial component than in the angular component which compromises the construction of a Poincaré map already for rather small noise levels. The variations in the angular component are typically unknown. Using an ensemble of potential Poincaré hyperplanes and employing statistical techniques might be a possible pathway to tackle such scenarios. In some rare situations it might be possible to measure the disturbance in the process period almost instantaneously. In such cases, a flexible sampling step depending on the observed disturbances in angular velocity could be used to damp the effect of the disturbance on the Poincaré map construction. Basically, an experimenter could either correct the sampling assuming an unperturbed constant period length or adjust it in accordance to a fixed Poincaré hyperplane in state space. However, the best procedure depends on the studied system. In our simulations the former approach should lead to a bias in the radial dynamics with linear changes in the control parameter $\mu$, whereas the first approach avoids that bias by an equidistant sampling of the changing radial dynamics.  

\subsection*{Data and Code Availability}
The $CaCO_3$ concentration time series from the Pacific sediment core at spot DSDP1218 is freely available under \url{https://www.ncei.noaa.gov/pub/data/paleo/contributions_by_author/tripati2005/tripati2005.txt}. The synthetic data of the methodological study in section \ref{sec: theory} can be found at \url{https://github.com/MartinHessler/Anticipation_of_Oligocenes_climate_heartbeat}. The methods used for the analysis are implemented in the open source python package \textit{antiCPy} which can be found at \url{https://github.com/MartinHessler/antiCPy} under a \textit{GNU General Public License v3.0}.

\subsection*{Author Contributions}
M.H. and O.K. designed research. M.H. implemented the method, performed data analysis and research and designed graphics. M.H. wrote the manuscript in consultation with O.K.

\subsection*{Additional Statements}
This research received no external funding. The authors declare no conflict of interest.

\subsection*{Acknowledgments}
M.H. thanks the Studienstiftung des deutschen Volkes for a scholarship including financial support.

\bibliographystyle{unsrt}
\bibliography{references.bib}

\newpage
\appendix

\section{Methods}
We shortly summarize the methods that we used to estimate a suitable embedding dimension in subsection \ref{subsec: fnn} and to control for the dependence on the time lags $\tau$ (in the case of the $CaCO_3$ concentration time series) in subsection \ref{subsec: add}.
\subsection{False next neighbour algorithm}
\label{subsec: fnn}
The \textit{false next neighbour} \textit{(fnn)} algorithm introduced in~\cite{a:kennel} counts the number of neighbouring states in an embedding that are no neighbours in an embedding of higher dimension. Therefore, a pair is classified as a neighbour pair if the distance of them is smaller than a predefined threshold value $R_{\text{threshold}}$. The neighbour pairs are found via the Python function \textit{sklearn.neighbors.NearestNeighbors}\cite{scikit-learn}. The ratio of false next neighbours can be determined by the formula~\cite{b:kantz}
\begin{align*} \label{Eq: fnn}
X_{\text{fnn}} = \frac{\sum_{n=1}^{N-m-1} \Theta \left( \frac{|s_n^{(m+1)}-s_{k(n)}^{(m+1)}|}{|s_n^{(m)}-s_{k(n)}^{(m)}|}-R_{\text{threshold}}\right)\Theta \left( \frac{\sigma}{R_{\text{threshold}}}-|s_n^{(m)}-s_{k(n)}^{(m)}|\right)}{\sum_{n=1}^{N-m-1} \Theta \left( \frac{\sigma}{R_{\text{threshold}}}-|s_n^{(m)}-s_{k(n)}^{(m)}|\right)}
\end{align*}
with the standard deviation $\sigma$, the next neighbour $s_{k(n)}^{(m)}$ of $s_n^{(m)}$ in $m$ dimensions and with $k(n)$ denoting the index $k$ with $|s_n-s_k| = \min$. The function $\Theta (\cdot )$ is the Heaviside function. The threshold has to be chosen large in order to take into account exponential divergence due to deterministic chaos. The first Heaviside function in the nominator yields unity if the neighbours in $m$ dimensions are false next neighbours. In other words, their distance ratio increases over the critical threshold $R_{\text{threshold}}$. The second Heaviside function filters all pairs that cannot be false next neighbours due to an initial distance greater than $\frac{\sigma}{R_{\text{threshold}}}$, because on average a displacement farther than $\sigma$ will not be realised. Depending on the threshold $R_{\text{threshold}}$ the ratio $X_{\text{fnn}}$ of false next neighbours should decrease with increasing embedding dimension $m$.
\subsection{Average distance from diagonal}
\label{subsec: add}
The mathematical Takens' theorem does not state anything about a convenient time lag $\tau$. A delay embedding in $m$ dimensions with two different time lags $\tau$ would be equivalent from the mathematical point of view. However, in practise it is necessary to take into account the relation between the internal dynamics' time scale and the sampling rate of a time series and in consequence the time lag $\tau$. If the time lag is small compared to the internal dynamics' time scale, the delay embedded values are very similar to each other. This strong correlation will lead to points located along the main diagonal of the delay embedding space. If the time lag is large compared to the internal dynamics' time scale, the correlation approaches zero and the points fill a large cloud in the $\mathbb{R}^m$ whereby the deterministic part is confined to very small scales of the structure.~\cite{b:kantz, a:rosenstein} \\
This geometrical reasoning motivates the method of \textit{average distance from diagonal}, abbreviated \textit{add}. The measure for the average distance $\langle S_m\rangle$ from diagonal for a given delay embedding is defined as 
\begin{align}
\langle S_m\left(\tau \right)\rangle =\frac{1}{N}\sum_{i=1}^N |\vec{X}_i^\tau -\vec{X}_i^0| 
\end{align} 
with the Euclidean distance measure $|\cdot |$ of the delay vector $\vec{X}_i^\tau$ from the non-delayed vector $\vec{X}_i^0$. In practise, the average distance from diagonal can be calculated with the scalar time series $X_t$ and the formula
\begin{align}
\langle S_m\left(\tau \right)\rangle =\frac{1}{N}\sum_{i=1}^N \sqrt{\sum_{j=1}^{m-1} (X_{i+j\tau}-X_i)^2}.
\end{align}
An optimal time lag $\tau$ is found for a maximum average distance $\langle S_m\left(\tau \right)\rangle$ from diagonal, because the attractor is well unfolded and the \textit{redundancy error} is inverse proportional to the average distance from diagonal and thus, minimized. The redundancy error is typically great for small delay times because successive coordinates are almost strongly correlated and similar. In this case additive noise has a relatively strong effect. The redundancy error decreases with bigger time lags $\tau$ and reaches a minimum because an attractor is bound to a certain volume in state space. The \textit{irrelevance error} is the error due to dynamical error of exponential growth. The irrelevance error increases with greater time lag because of the extended time for this exponential divergence with greater time lag $\tau$. In reality it would be necessary to find the best trade-off between redundancy error and irrelevance error, but quantifying the latter is not possible. For this reason the effects of irrelevance error are assumed to be small and negligible relative to the redundancy error.~\cite{a:rosenstein}
\section{Additional Information: Synthetic Data}
In this section we summarize important information about the synthetic data of the iterated maps in subsection \ref{subsec: appendix synthetic} and of the limit cycle model in subsection \ref{subsec: appendix limit cycles}.~\cite{mt:Hessler2019}
\subsection{Iterated maps}\label{subsec: appendix synthetic}
\begin{figure*}[!htbp]
    \subfigure[Optimal dimension $d_{\rm opt} \approx 3$]{
    \label{subfig: fnn nm}
    \includegraphics[width=0.32\textwidth]{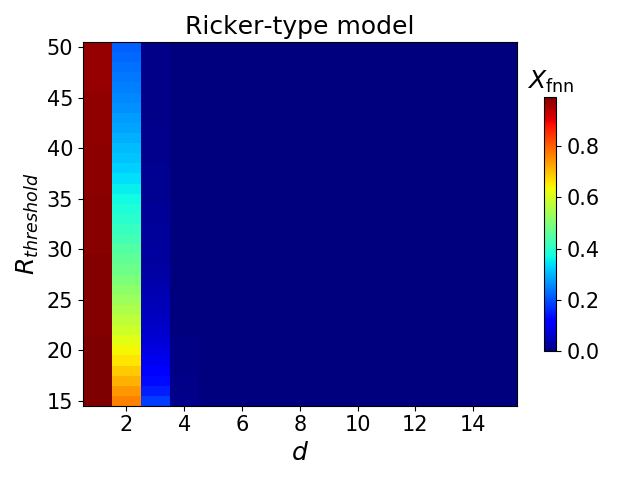}}
    \subfigure[Optimal dimension $d_{\rm opt} \approx 3$]{
    \label{subfig: fnn henon}
    \includegraphics[width=0.32\textwidth]{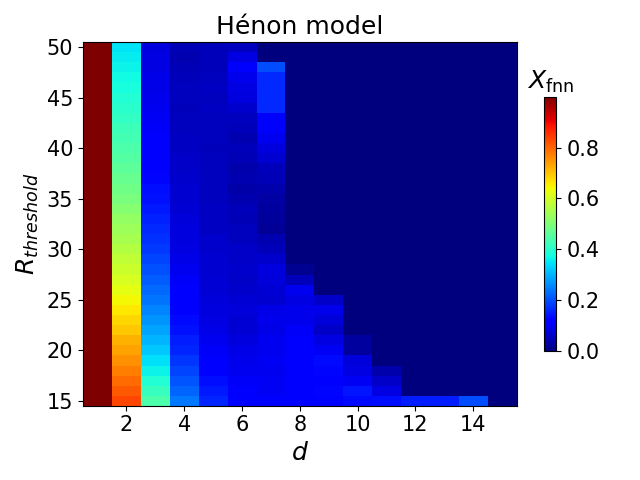}}
    \subfigure[Optimal dimension $d_{\rm opt} \approx 4$]{
    \label{subfig: fnn hopf}
    \includegraphics[width=0.32\textwidth]{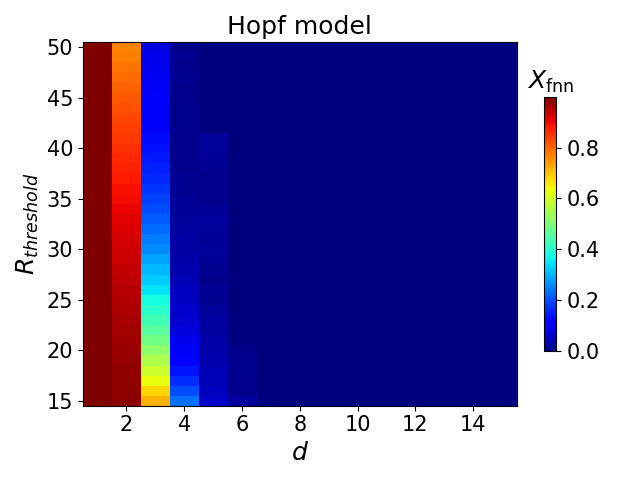}}
\caption[False next neighbour results for the Ricker-type, the Hénon and the Hopf model.]{The false next neighbour algorithm is applied over a wide range of threshold values $R_{\text{threshold}}$ to ensure independence of the results from the chosen classification threshold. The determined optimal embedding dimensions coincide with the optimal quality of bifurcation classification in the complex plane in subsection \ref{subsec: quality results}.}
\label{fig: fnn synthetic data}
\end{figure*}

The false next neighbour algorithm is applied to the investigated datasets of the sections \ref{subsec: datasets} over a range of threshold values to ensure robustness of the results which are shown in figure \ref{fig: fnn synthetic data}.

\subsection{Limit cycles}\label{subsec: appendix limit cycles}
\begin{figure*}[!htbp]
    \subfigure[Model with $\sigma_r = 0.01$.]{
    \label{subfig: fnn 1noise001}
    \includegraphics[width=0.48\textwidth]{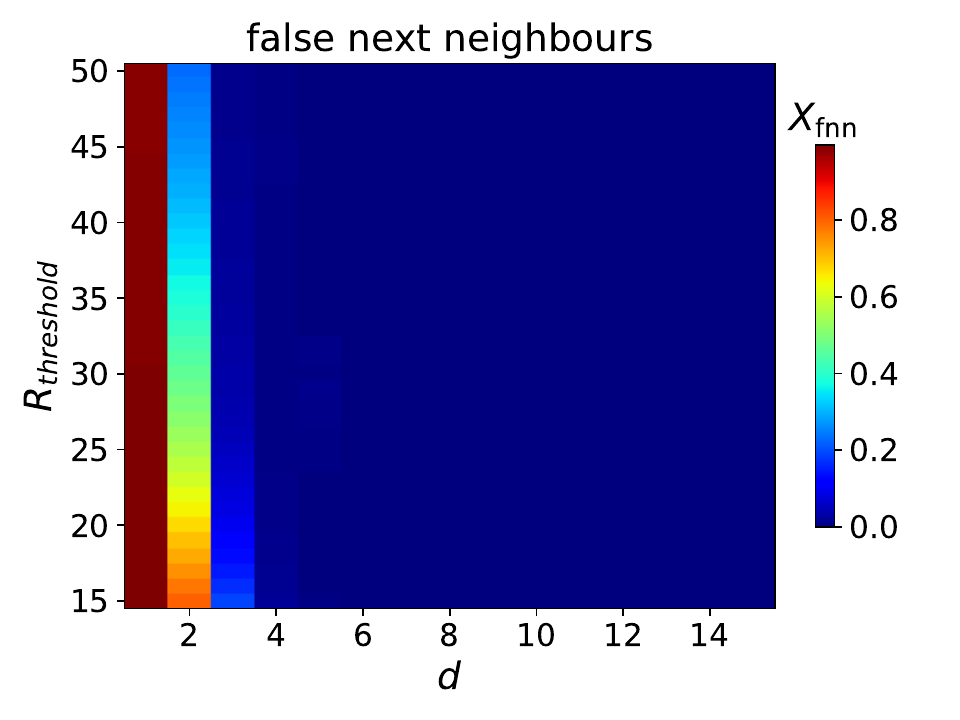}}
    \subfigure[Model with $\sigma_r = 0.05$.]{
    \label{subfig: fnn 1noise005}
    \includegraphics[width=0.48\textwidth]{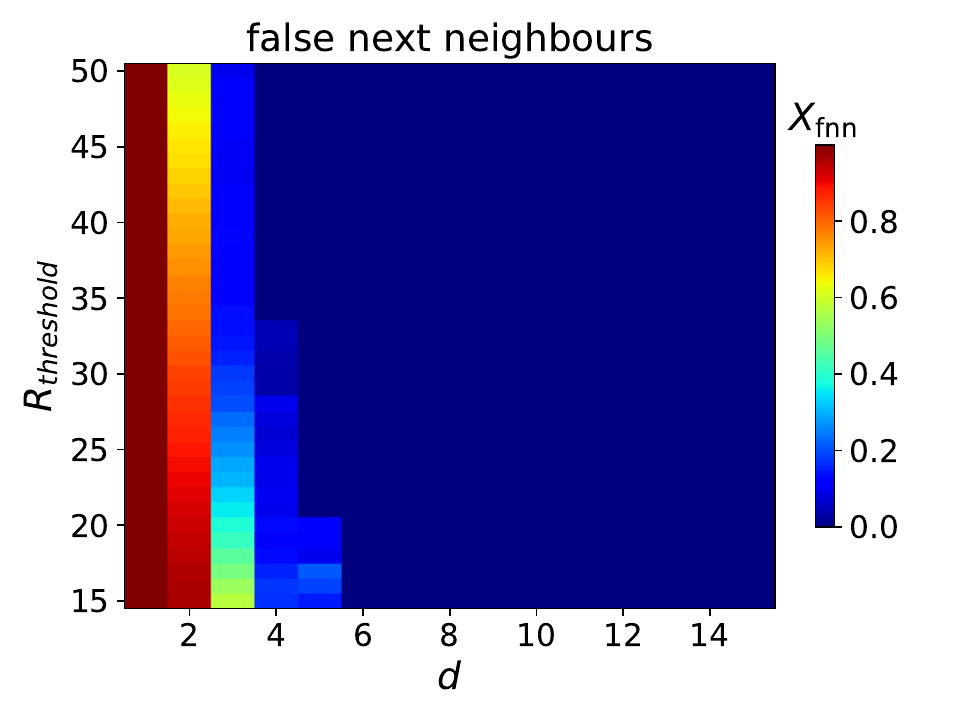}}
    \subfigure[Model with $\sigma_r = 0.01$ and $\zeta = 0.1$.]{
    \label{subfig: fnn 2noise001scale01}
    \includegraphics[width=0.48\textwidth]{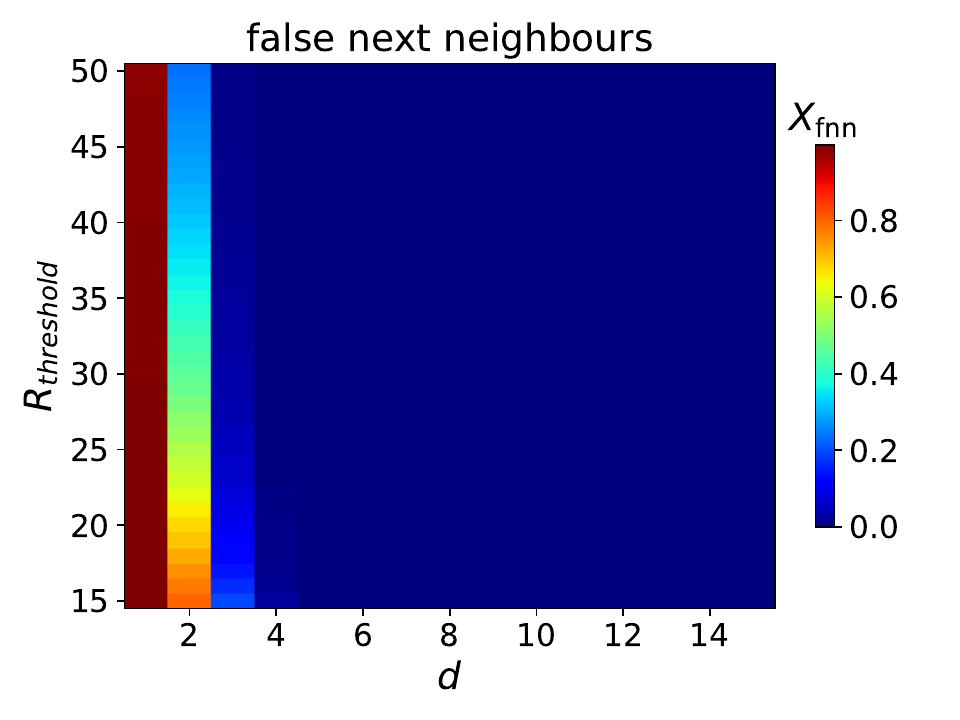}}
        \subfigure[Model with $\sigma_r = 0.01$ with $\zeta = 0.5$.]{
    \label{subfig: fnn 2noise001scale05}
    \includegraphics[width=0.48\textwidth]{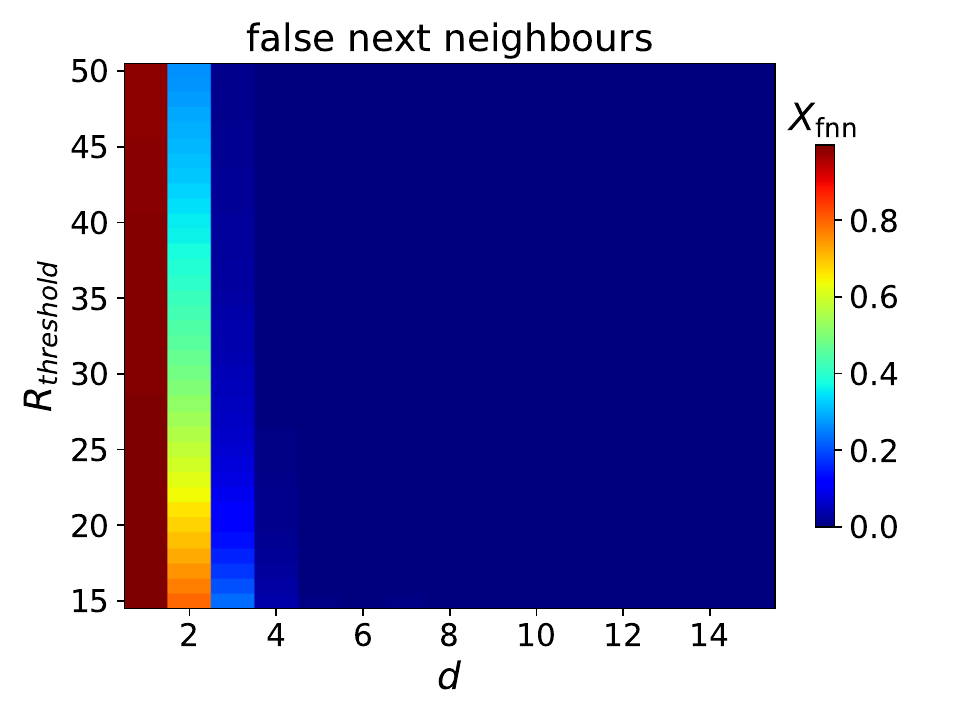}}
\caption[False next neighbour results for the limit cycle stability analysis.]{The false next neighbour algorithm is applied over a wide range of threshold values $R_{\text{threshold}}$ to the four Poincaré maps that are constructed from the limit cycle model eq. \ref{eq: limit cycle}. the fnn results for the models are as follows: (a) $\sigma_r = 0.01$, (b) $\sigma_r = 0.05$, (c) $\sigma_r = 0.01$ with $\zeta = 0.1$, (d) $\sigma_r = 0.01$ with $\zeta = 0.5$. In general, the fnn algorithm suggests an optimal embedding dimension $d_{\rm opt} \approx 3$. In our analysis we choose a slightly smaller embedding dimension $d=2$, since the results are comparable and provide stable estimation of the Gaussian plane for even smaller window sizes $n_{\rm w}$ $n_{\rm w}$ as it would be possible for embedding dimension $d_{\rm opt} \approx 3$.}
\label{fig: fnn limit cycle}
\end{figure*}

We present the results of the fnn algorithm applied to the four Poincaré maps of our analysis in subsection \ref{subsec: results limit cycles} over a wide range of threshold values $R_{\text{threshold}}$ in figure \ref{fig: fnn limit cycle}. In the analysis of the main article we use an embedding dimension $d=2$ that is slightly smaller than the optimal embedding dimension $d_{\rm opt} \approx 3$, because the DEV results fluctuate even less and are more stable for smaller window sizes $n_{\rm w}$ if we used the embedding dimension $d=2\lesssim d_{\rm opt} \approx 3$. However, the proposed fnn optimal embedding dimension yields qualitatively similar results. The chosen method parameters of the DEV are listed in table  \ref{table: window params}.\\
The drift slopes are derived from the drift parameterisation
\begin{align}
    D_1(x)=\theta_0+\theta_1\cdot x+\theta_2\cdot x^2 + \theta_3\cdot x^3
    \label{eq:langevin}
\end{align}
of a drift diffusion model with constant diffusion $D_2(x,t)\equiv const.$ with a Standard Wiener process. the parameters $\theta_i$ are sampled via a Bayesian Markov Chain Monte Carlo approach. The prior sampling ranges of the parameters $\theta_i$ are listed in table \ref{table: resilience priors}. More details about the drift slope estimation can be found in Heßler and Kamps~\cite{a:Hessler2021, a:Hessler2022}.

\renewcommand{\arraystretch}{1.1}
\begin{table*}[h]
\centering
\begin{tabular}{@{\extracolsep{4pt}}lcccccc}\hlinewd{1pt}
\textbf{Estimate} & \multicolumn{2}{c}{\textbf{Rolling Window}} & \multicolumn{4}{c}{\textbf{Method Parameters}}\\
\cline{2-3} \cline{4-7}\\
&\textbf{Size}&\textbf{Shift}&\textbf{Dimension}& \textbf{Walkers}&\textbf{Steps}&\textbf{Burn In}\\\hline
DEV $\hat{\lambda}_{\rm max}$ &2000&1&2&---&---&---\\
Drift Slope $\hat{\gamma}$&3000&25&---&50&15000&200\\
\hlinewd{1pt}
\end{tabular}
\caption{Parameters that are used to compute the DEV and the drift slope estimates for the four limit cycle examples via an on-line approach.}
\label{table: window params}
\end{table*}

\begin{table}
\centering
\begin{tabular}{ccccc}\hlinewd{1pt}
\multicolumn{5}{c}{\textbf{Prior (Range)}}\\
\hline\\
$\boldsymbol\theta_0$&$\boldsymbol\theta_1$&$\boldsymbol\theta_2$&$\boldsymbol\theta_3$&$\boldsymbol\theta_4$\\\hline
[-50,50] & [-50,50] & [-50,50] & [-50,50] & [0,50]\\
\hlinewd{1pt}
\end{tabular}
\caption{Prior configurations of the parameters of eq. \ref{eq:langevin}.}
\label{table: resilience priors}
\end{table}
\section{Additional information: greenhouse-icehouse transition data}
\label{sec: appendix greenhouse}
In the following sections we sketch the preprocessing of the $CaCO_3$ concentration and the parameter optimisation. The procedures and results are summarized in figure \ref{fig: greenhouse preparation}. Furthermore, we perform a detailed robustness check against other parameter choices and discuss briefly the delay embedding of the data in three dimensions with and without detrending.
\subsection{Data interpolation}
\label{subsec: appendix data preparation}
We focus our data analysis on the period prior to the greenhouse-icehouse transition. The data obtained from the sediment cores are commonly not equidistantly sampled and furthermore the studied time series contains duplicates for certain time stamps. We treat these duplicates by replacing them by their averages. Afterwards we perform a linear interpolation of the data maintaining the original number of measurement points ($n=461$). The interpolation is performed via \textit{scipy.interpolate.interp1d}\cite{2020SciPy-NMeth}. The treatment is almost identical to that of previous work~\cite{a:dakos08}. The interpolation procedure is illustrated in subfigure \ref{subfig: interpolation}.
\subsection{Gaussian Detrending}
\label{subsec: appendix gauss detrend}
\begin{figure*}
    \centering
\subfigure[Data Interpolation]{\label{subfig: interpolation}\includegraphics[width=0.31\textwidth]{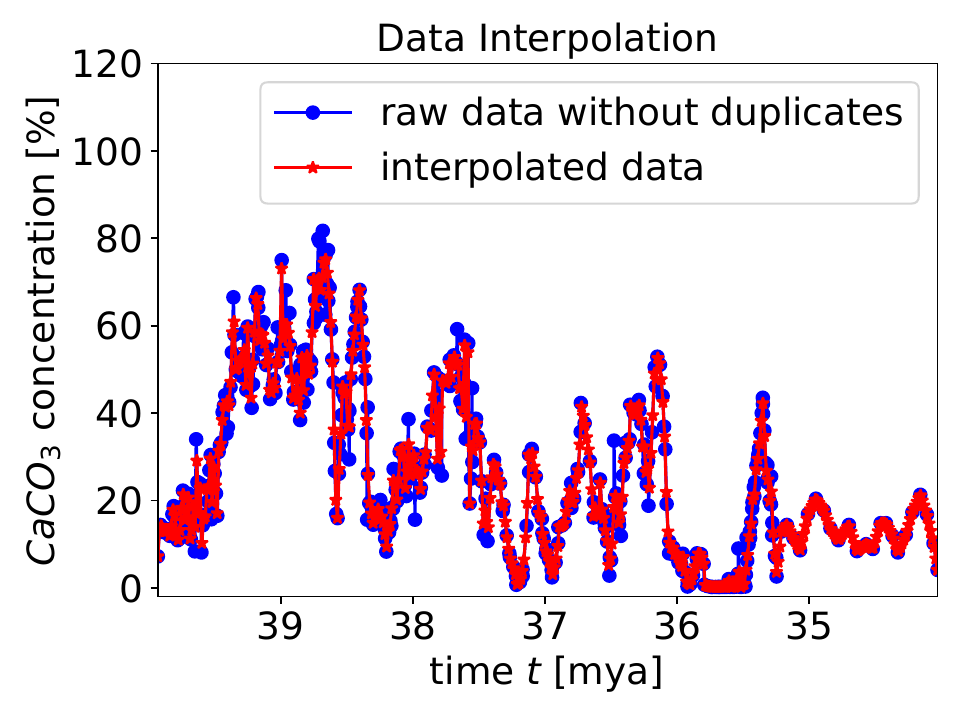}}
\addtocounter{subfigure}{1}
\subfigure[fnn-Results of raw data.]{
\label{subfig: fnn raw greenhouse}
\includegraphics[width=0.32 \textwidth]{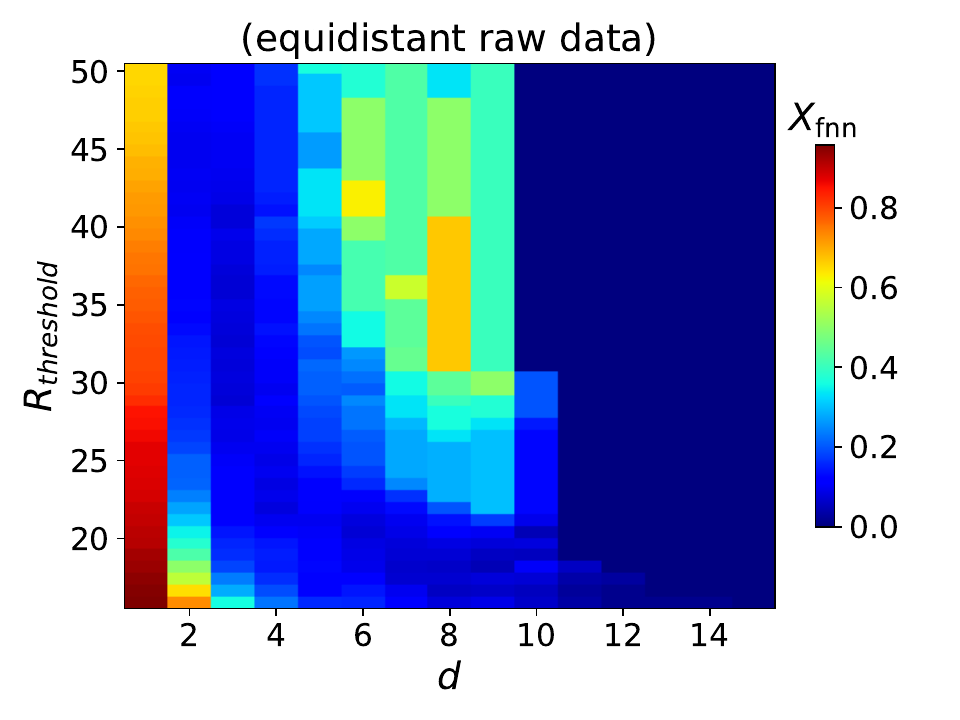}}
\addtocounter{subfigure}{1}
\subfigure[add-Results of raw data.]{\label{subfig: time lag}\includegraphics[width= 0.32\textwidth]{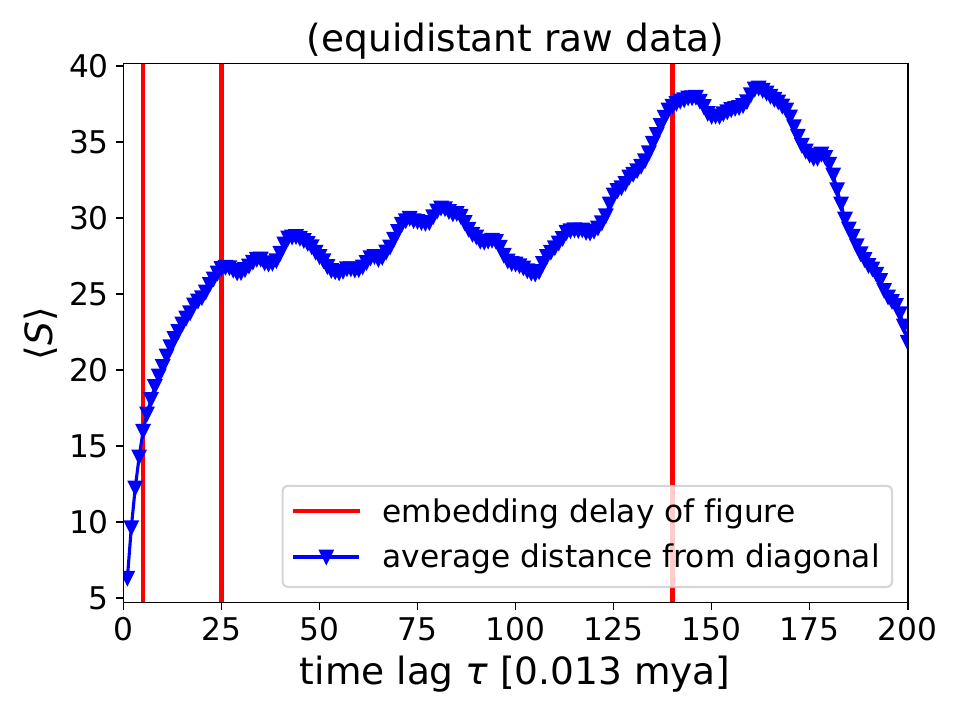}}
\addtocounter{subfigure}{-4}
\subfigure[Gaussian Detrending.]{\label{subfig: greenhouse detrending}\includegraphics[width= 0.31\textwidth]{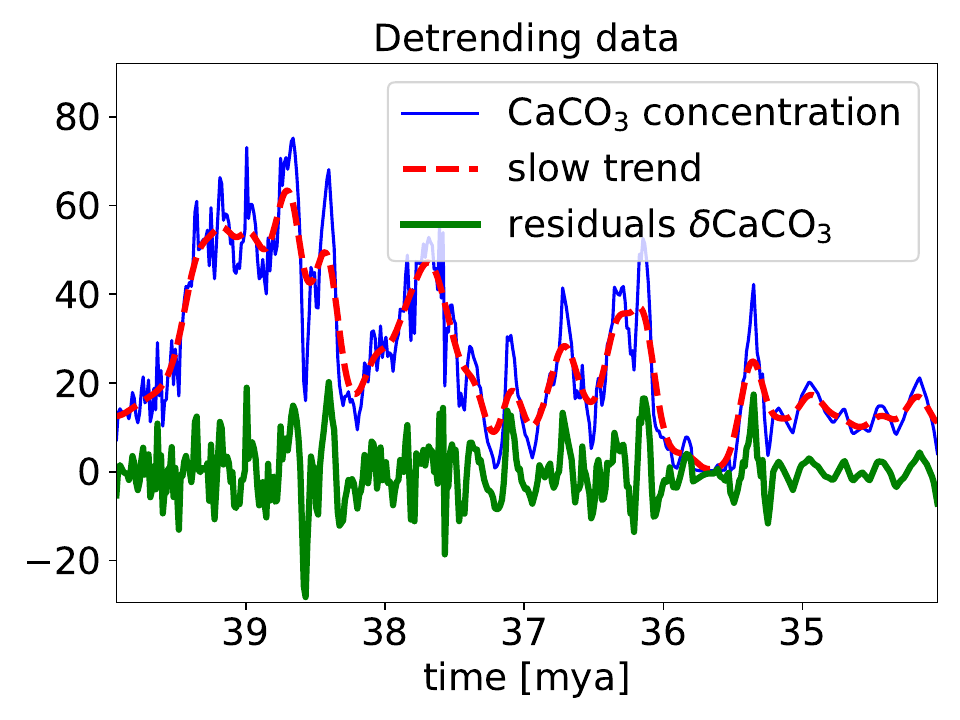}}
\addtocounter{subfigure}{1}
\subfigure[fnn-Results of detrended data.]{
\label{subfig: fnn detrended greenhouse}
\includegraphics[width=0.32\textwidth]{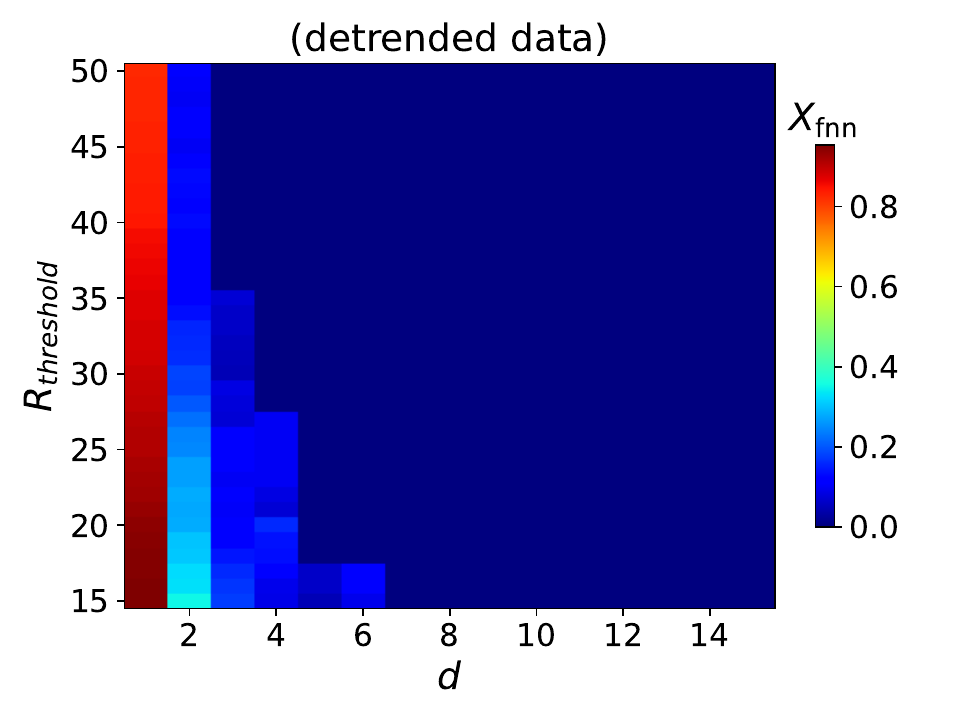}}
\addtocounter{subfigure}{1}
\subfigure[add-Results of detrended data.]{\label{subfig: time lag detrended}\includegraphics[width= 0.32\textwidth]{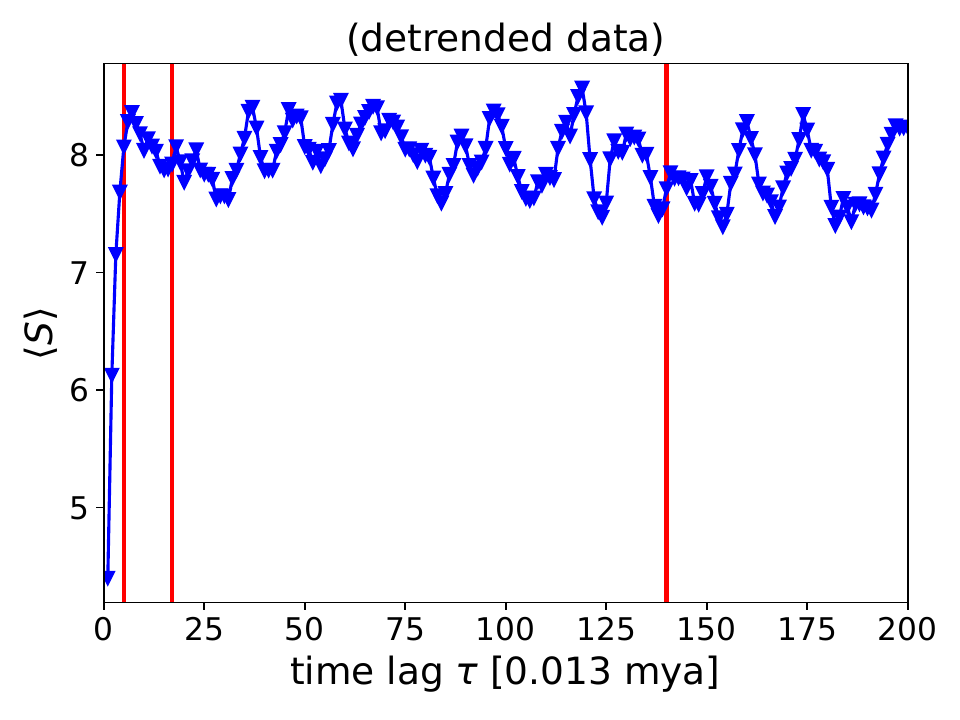}}
\caption[Preparation of the greenhouse-icehouse transition analysis.]{(a) Analogue to~\cite{a:dakos08} we perform an interpolation of the data after replacing doubled values for time stamps by their average. The number of interpolated points corresponds to the number of original data points, i.e. $n=461$ in our analysis. (b) Following\cite{a:dakos08}, a Gaussian kernel is used to fit a slow trend in the $CaCO_3$ concentration data. The trend is subtracted from the raw data to obtain the stationary detrended versions for the analysis. We show the slow trend fit with a Gaussian kernel bandwidth of $\tilde{\sigma}=5$ as an example in subfigure \ref{subfig: greenhouse detrending}. (c,d) In subfigure \ref{subfig: fnn raw greenhouse} we show the false next neighbour results for the original $CaCO_3$ concentration time series over a $R_{\rm threshold}$-range. Since we analyse detrended versions of the dataset, we present the fnn analysis for the detrended data of subfigure \ref{subfig: greenhouse detrending} in subfigure \ref{subfig: fnn detrended greenhouse}. Both fnn analyses suggest a proper embedding dimension from roughly $d\sim\SIrange{3}{5}{}$. (e,f) In order to give at least some viusalizations of possible embeddings we also perform a add-analysis for the three dimensional case for both the raw and detrended data in subfigures \ref{subfig: time lag} and \ref{subfig: time lag detrended}, respectively. We chose three time lags corresponding to the red vertical lines to construct the shadow manifolds in figure \ref{fig: shadow attractors analysed period} of section \ref{subsec: shadow manifold}.
}
\label{fig: greenhouse preparation}
\end{figure*}

In contrast to the synthetic datasets that we analysed in the main article, a simple linear detrending per window does not seem to comply with the complex slow trend behaviour of the greenhouse-icehouse transition time series. Therefore, we follow an analogue approach to Dakos et al.\cite{a:dakos08} that we illustrate in figure \ref{fig: greenhouse preparation}. We subtract a Gaussian kernel smoothed fit from the raw data to guarantee stationarity. The kernel smoothing is performed using \textit{scipy.ndimage.filters}\cite{2020SciPy-NMeth}. The detrending is performed for various kernelwidth $\tilde{\sigma} = \lbrace 5,10,20\rbrace$ and truncated at $\tilde{\sigma} = 4$. An example result with a kernelwidth of $\tilde{\sigma}=5$ is given in figure \ref{subfig: greenhouse detrending}.
\subsection{False next neighbour results}
\label{subsec: appendix fnn greenhouse}
The fnn algorithm (cf. Appendix \ref{subsec: fnn}) is applied to the $CaCO_3$ concentration data over a range of threshold values $R_{\text{threshold}}$. The optimal embedding dimension is $d_{\rm opt} \approx\SIrange[]{3}{5}{}$. We choose $d_{\rm opt}=5$ in analogy to Grziwotz et al.~\cite{a:grziwotz}.
The results are shown in subfigures \ref{subfig: fnn raw greenhouse} and \ref{subfig: fnn detrended greenhouse}.
\subsection{Average distance from diagonal results}
The average distance from diagonal (add) is not considered for the actual DEV analysis, since the AR model by definition does not depend on a proper embedding as e.g. the S-map approach used in~\cite{a:grziwotz}. The AR estimation would basically loose information if we thin the data, because its coefficients are fitted to predict future points based on past data points and not as for the S-map approach by a cloud of next neighbours in state space.\\ 
Here we compute the optimal time delay $\tau$ in units of sampling steps for an embedding dimension $p=3$ in order to visualize some unfoldings of the shadow manifold in subsection \ref{subsec: shadow manifold}. The results for the raw and detrended data examples are shown in subfigures \ref{subfig: time lag} and \ref{subfig: time lag detrended}, respectively.
\subsection{Statistical leading indicators}
\label{subsec: appendix std greenhouse}
\begin{figure}
    \centering
    \includegraphics[width=0.49\textwidth]{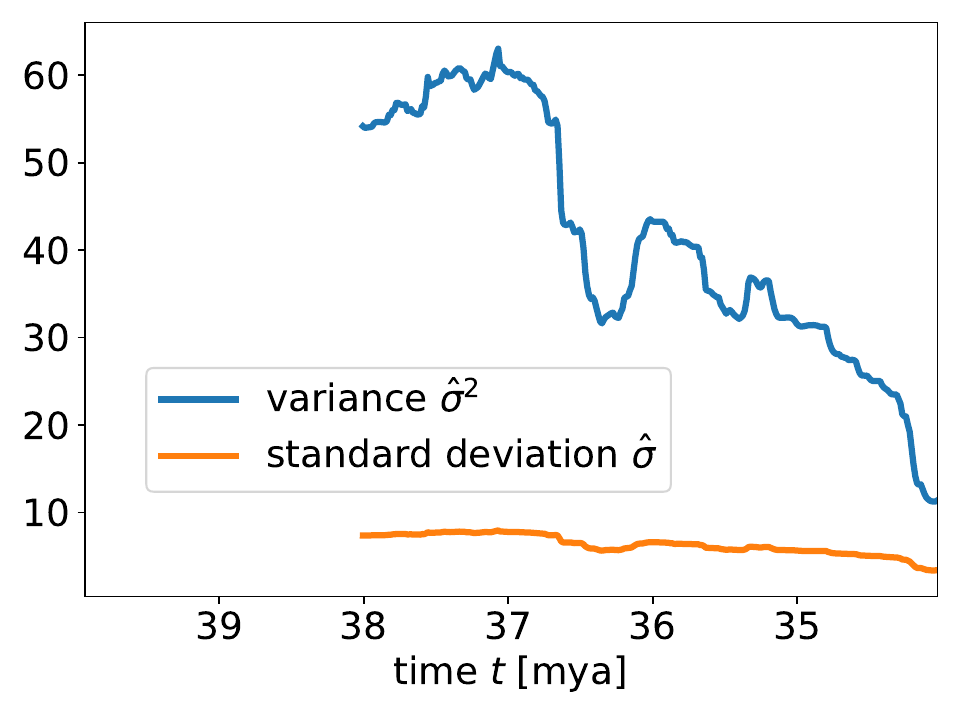}
    \caption{The variance of the considered data period exhibits a negative trend prior to the assumed bifurcation-induced transition from a greenhouse into an icehouse-earth state. In this sense, there is no evidence for critical slowing down given by the simple statistical measures, autocorrelation at lag-1 and variance which have to increase both at the same time to be early warning signals of an uprising bifurcation-induced transition.}
    \label{fig: variance greenhouse}
\end{figure}

As reported by Dakos et al.\cite{a:dakos08} the analysed time series exhibits a positive autocorrelation trend for lag-1 correlations. However, we cannot be sure about the presence of critical slowing down prior to a bifurcation, because the findings are not accompanied by a simultaneous increase in variance which would be expected in order to interpret the statistical measures \textit{combined} as an early warning sign. We show the variance and standard deviation of the analysed greenhouse-icehouse transition period in figure \ref{fig: variance greenhouse}.
\subsection{Visualization of shadow attractors}
\label{subsec: shadow manifold}
\begin{figure*} 
    \centering
\subfigure[$d=3\lesssim d_{\rm opt}$, lag $\tau=5$.]{\label{subfig: nd tau5}\includegraphics[width=0.31\textwidth]{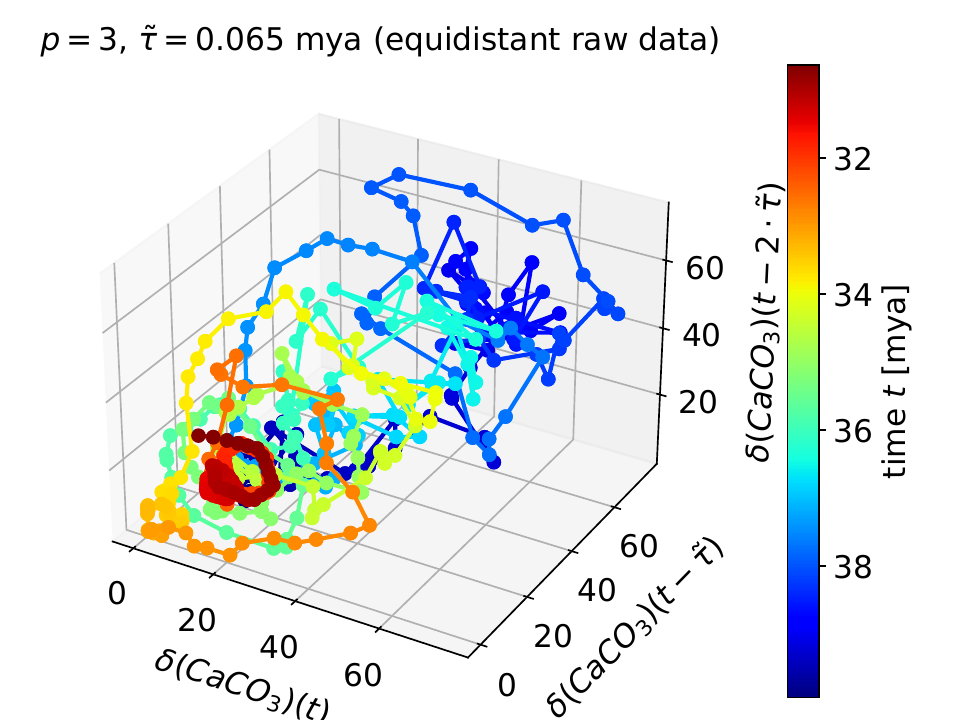}}
\addtocounter{subfigure}{1}
\subfigure[$d=3\lesssim d_{\rm opt}$, lag $\tau=25$.]{
\label{subfig:  nd tau25}
\includegraphics[width=0.32 \textwidth]{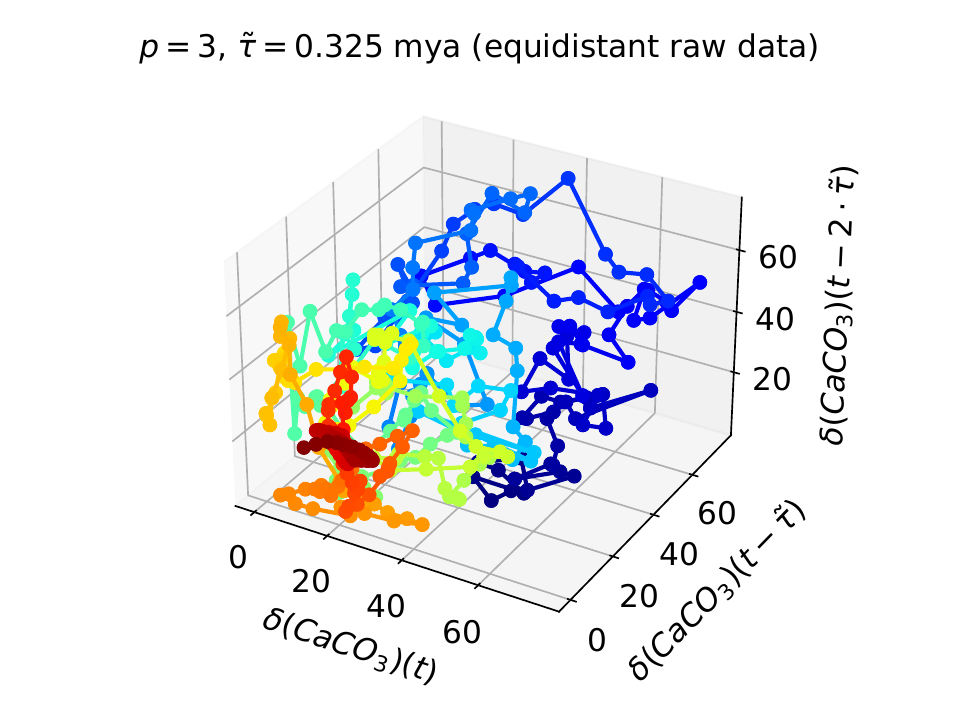}}
\addtocounter{subfigure}{1}
\subfigure[$d=3\lesssim d_{\rm opt}$, lag $\tau=140$.]{\label{subfig: nd tau140}\includegraphics[width= 0.32\textwidth]{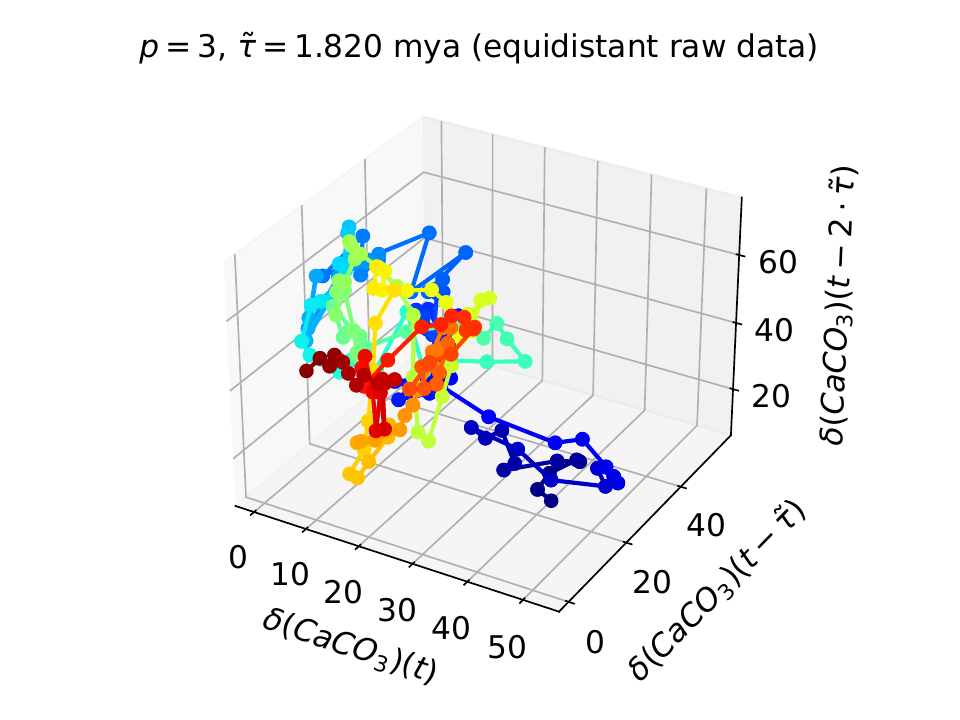}}
\addtocounter{subfigure}{-4}
\subfigure[$d=3\lesssim d_{\rm opt}$, lag $\tau=5$.]{\label{subfig: d tau5}\includegraphics[width= 0.31\textwidth]{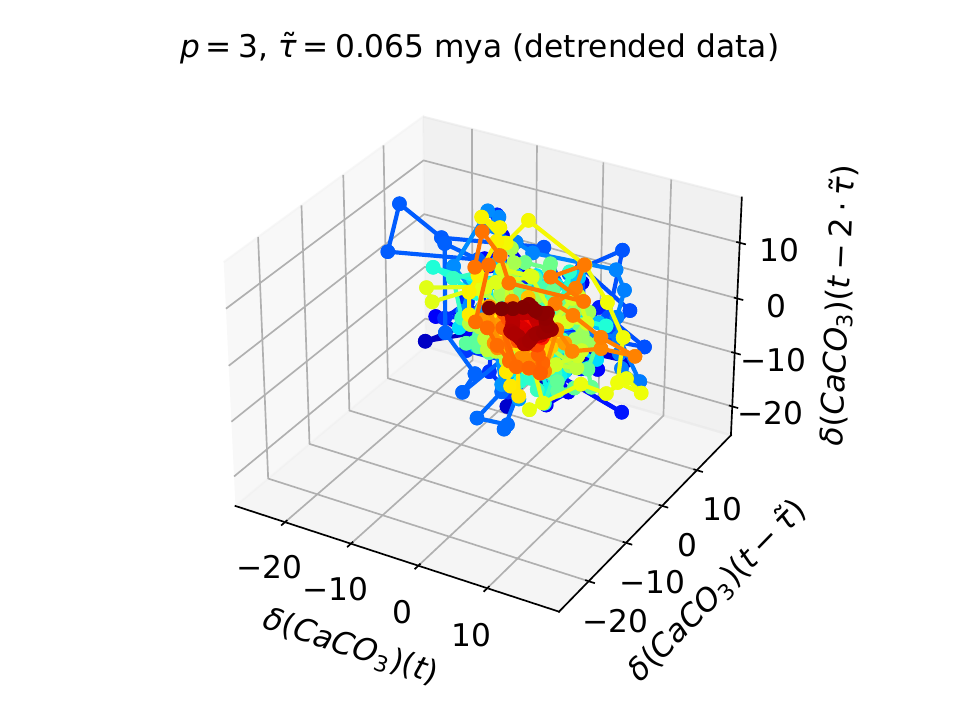}}
\addtocounter{subfigure}{1}
\subfigure[$d=3\lesssim d_{\rm opt}$, lag $\tau=17$.]{
\label{subfig:  d tau17}
\includegraphics[width=0.32\textwidth]{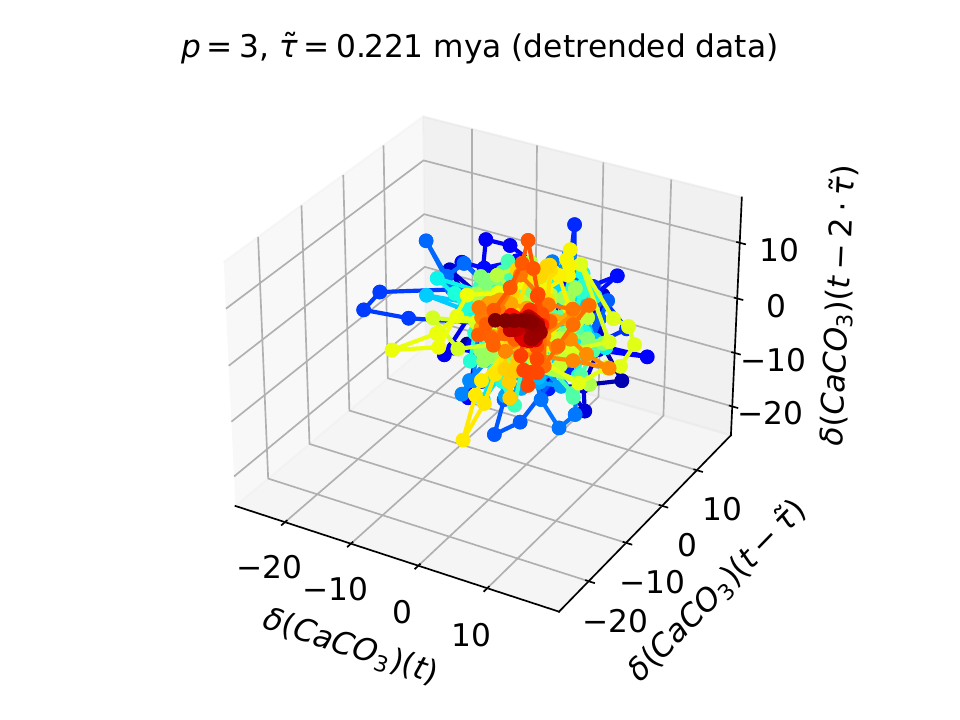}}
\addtocounter{subfigure}{1}
\subfigure[$d=3\lesssim d_{\rm opt}$, lag $\tau=140$.]{\label{subfig: d tau140}\includegraphics[width= 0.32\textwidth]{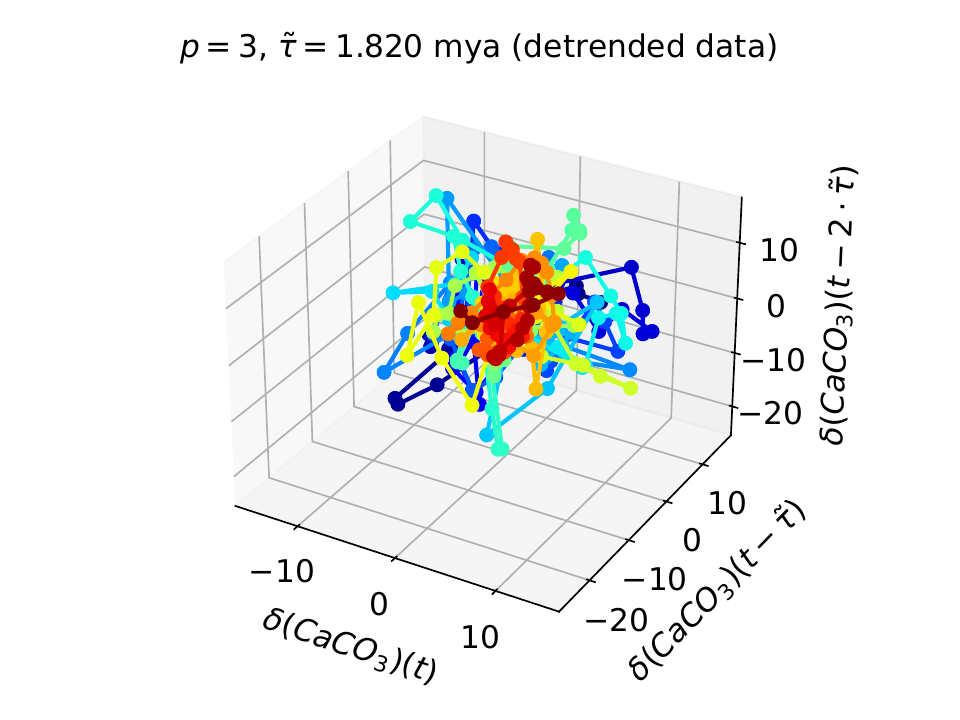}}
\caption[Exemplary delay embeddings of the analyses greenhouse-icehouse  transition time series period without and with detrending.]{We show unfoldings of the equidistant raw data and the detrended versions, respectively, for some example time delays $\tau$ which are chosen accordingly to the add results in subfigures \ref{subfig: time lag} and \ref{subfig: time lag detrended}.
}
\label{fig: shadow attractors analysed period}
\end{figure*}

\begin{figure*}
    \centering
\subfigure[$d=3\lesssim d_{\rm opt}$, lag $\tau=5$.]{\label{subfig: whole nd tau5}\includegraphics[width=0.31\textwidth]{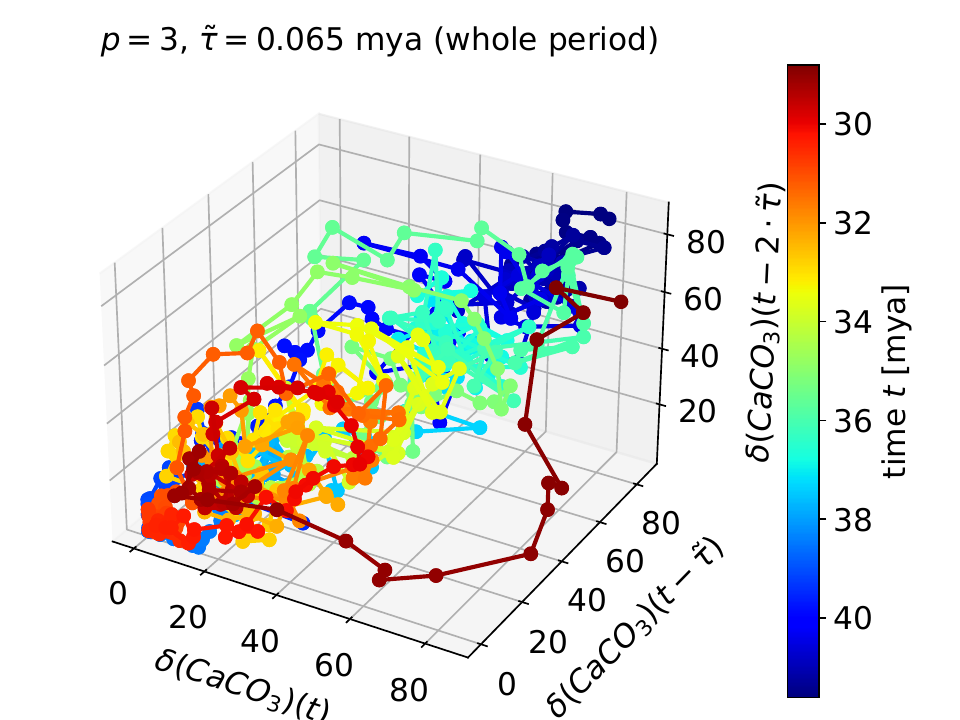}}
\addtocounter{subfigure}{1}
\subfigure[$d=3\lesssim d_{\rm opt}$, lag $\tau=57$.]{
\label{subfig:  nd tau57}
\includegraphics[width=0.32 \textwidth]{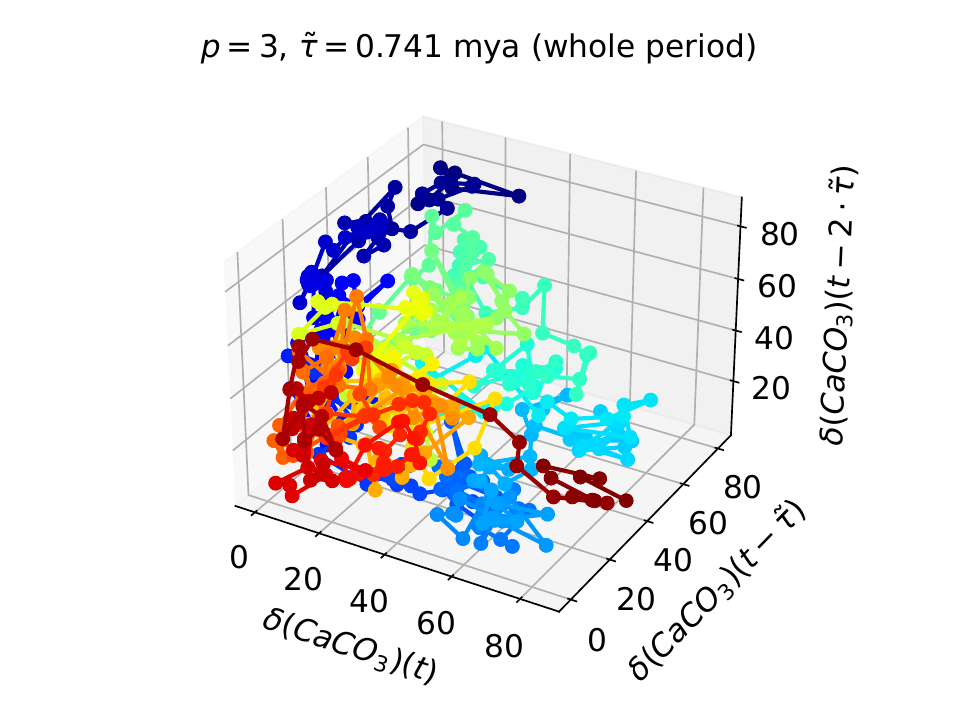}}
\addtocounter{subfigure}{1}
\subfigure[$d=3\lesssim d_{\rm opt}$, lag $\tau=140$.]{\label{subfig: whole nd tau140}\includegraphics[width= 0.32\textwidth]{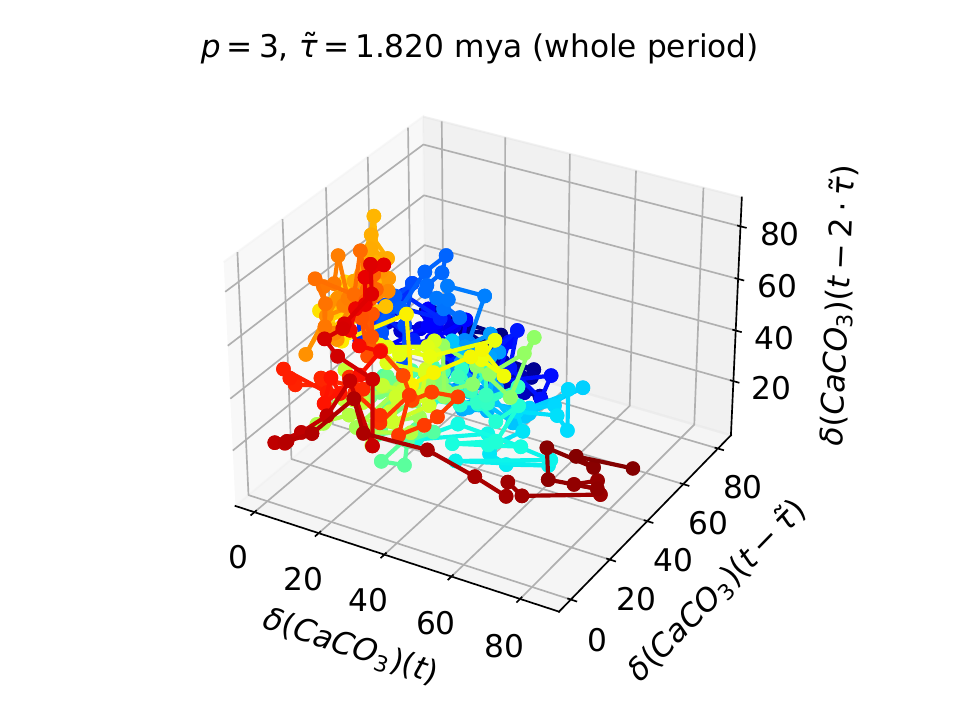}}
\addtocounter{subfigure}{-4}
\subfigure[$d=3\lesssim d_{\rm opt}$, lag $\tau=5$.]{\label{subfig: whole d tau5}\includegraphics[width= 0.31\textwidth]{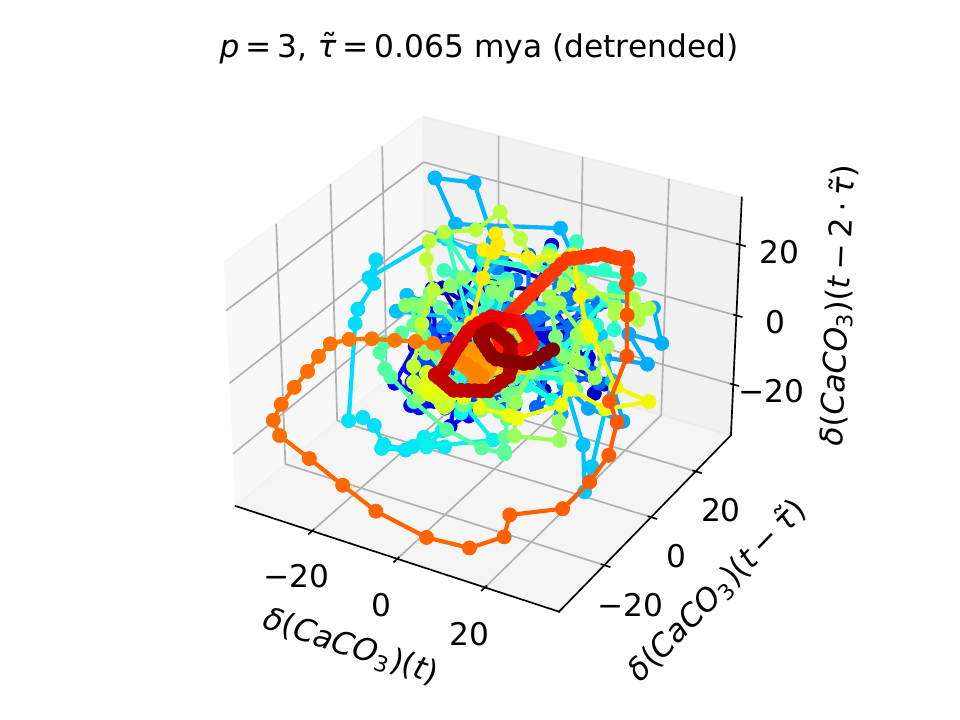}}
\addtocounter{subfigure}{1}
\subfigure[$d=3\lesssim d_{\rm opt}$, lag $\tau=25$.]{
\label{subfig: whole d tau25}
\includegraphics[width=0.32\textwidth]{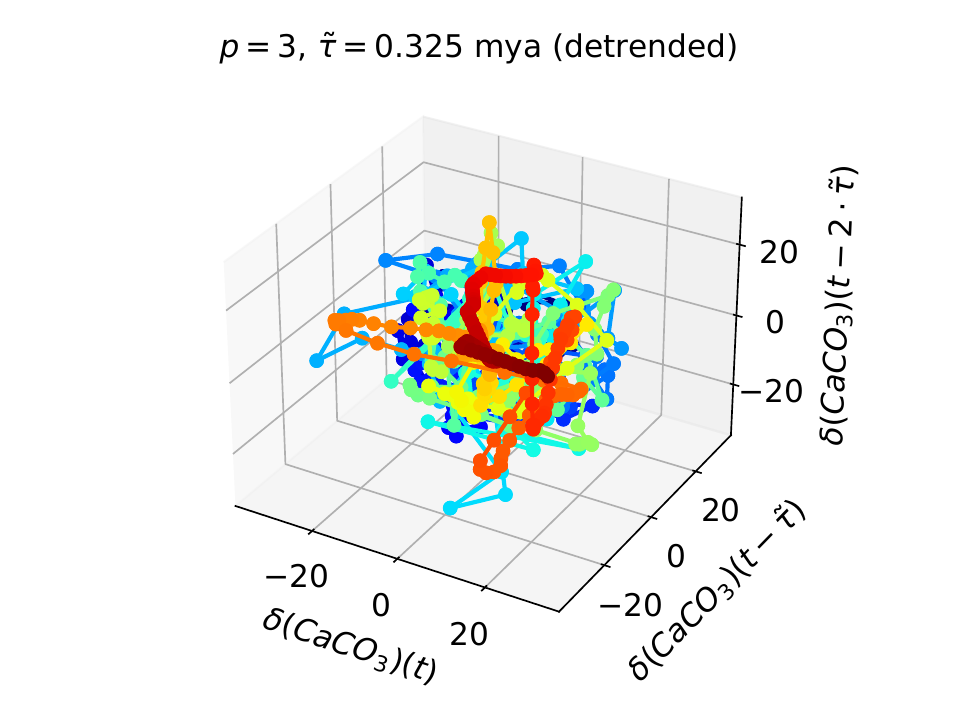}}
\addtocounter{subfigure}{1}
\subfigure[$d=3\lesssim d_{\rm opt}$, lag $\tau=123$.]{\label{subfig: whole d tau123}\includegraphics[width= 0.32\textwidth]{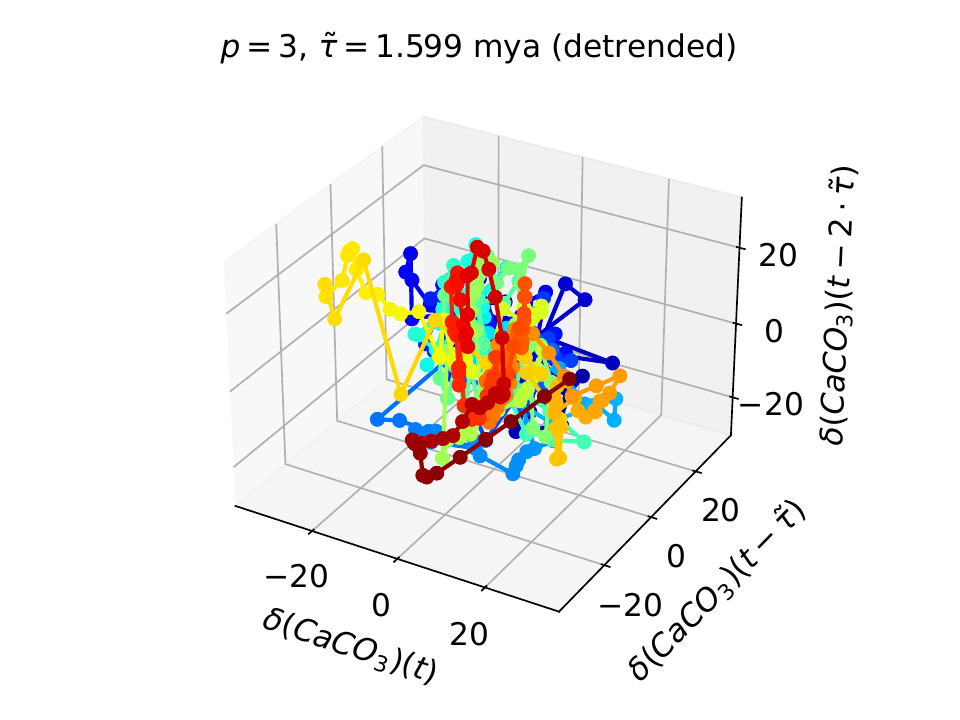}}
\caption[Exemplary delay embeddings of the greenhouse-icehouse transition time series's whole period without and with detrending.]{Same as in figure \ref{fig: shadow attractors analysed period}, but for the whole dataset. The add results are shown in figure \ref{fig: delay whole set}.
}
\label{fig: shadow attractors whole}
\end{figure*}

\begin{figure*}
    \centering
\subfigure{\label{subfig: whole delay}\includegraphics[width= 0.49\textwidth]{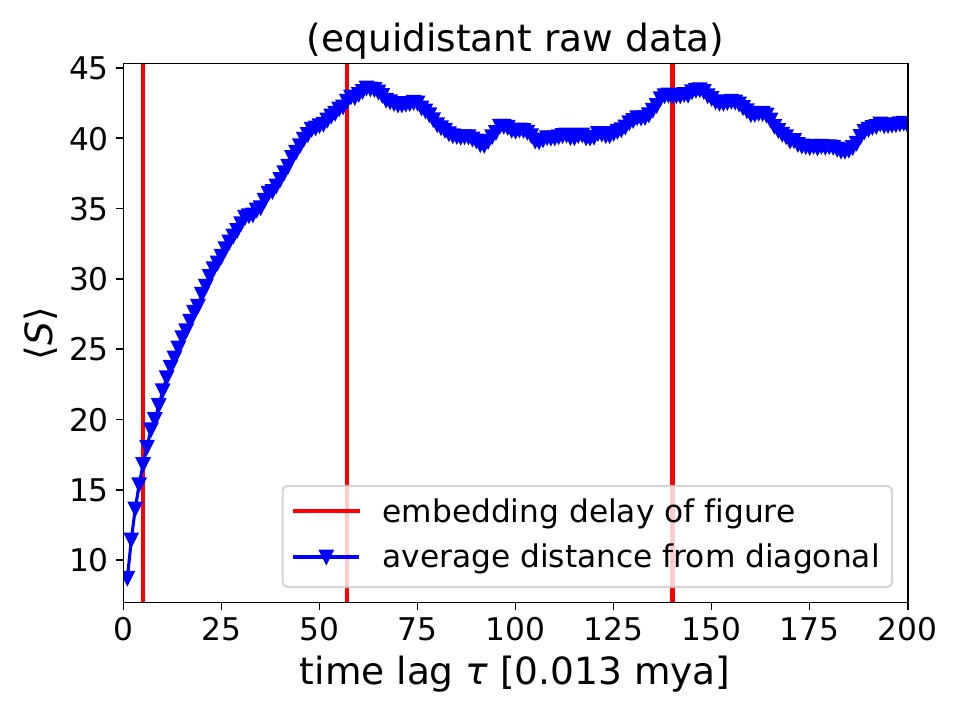}}
\subfigure[Detrended whole dataset.]{\label{subfig: detrended whole delay}\includegraphics[width= 0.49\textwidth]{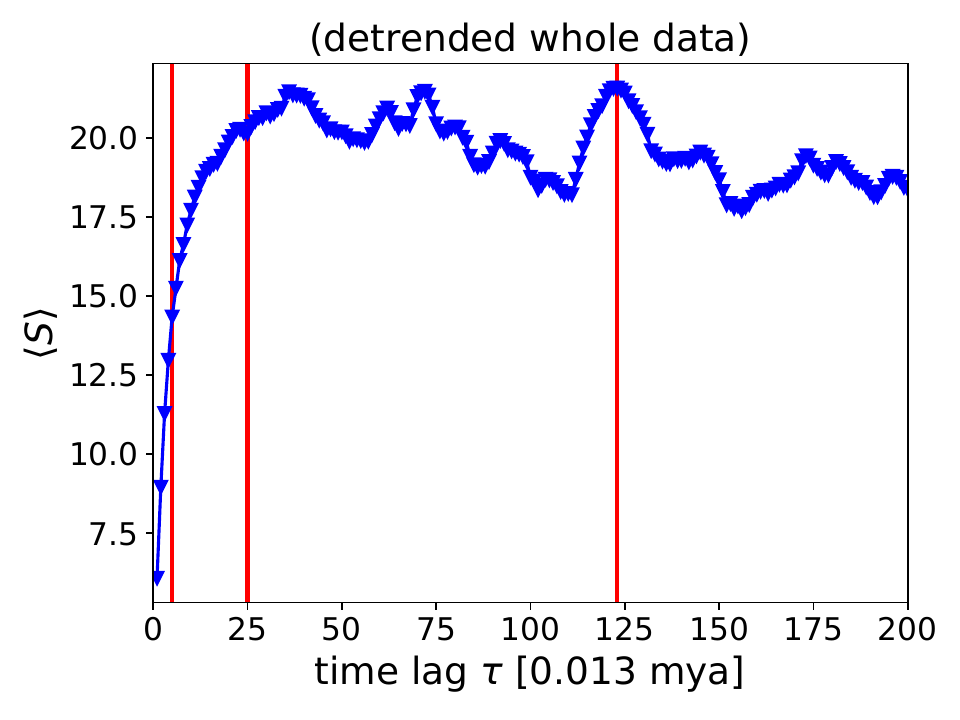}}
\caption[add algorithm for the whole greenhouse-icehouse transition data.]{Results of the add algorithm for the whole greenhouse-icehouse transition data without and with detrending. In this case we show the detrended version with kernelwidth $\tilde{\sigma} = 20$.
}
\label{fig: delay whole set}
\end{figure*}

We accompany the analysis by exemplary time delay embeddings in three dimensions. In contrast to the optimal delay embedding dimension $d_{\rm opt} = 5$ chosen in the main article, we fall back to an sub-optimal embedding dimension of $d=3\lesssim d_{\rm opt}=5$ to study the resulting time delay embeddings by eye (cf. fnn results in subfigures \ref{subfig: fnn raw greenhouse} and \ref{subfig: fnn detrended greenhouse}). In figure  \ref{fig: shadow attractors analysed period} we consider delay embeddings of data in the time interval from $\SIrange{34.03}{39.93}{mya}$. We use time delays from the add analysis shown in the subfigures \ref{subfig: time lag} and \ref{subfig: time lag detrended} for the equidistant raw data and the detrended data, respectively. The embedded shadow attractor (or since we do not know the exact mathematical object, something like a shadow trajectory) in subfigure \ref{subfig: nd tau5} exhibits a significant redundancy error~\cite{a:rosenstein}, i.e. the data is aligned along the main diagonal in three dimensions. We have to keep that in mind in order to interpret these imperfect visualizations carefully. However, we can already make an interesting observation in subfigure \ref{subfig: nd tau5}. Apart from a drift of the trajectory approximately from the upper right to the lower left over time (from blue to red), the embedding shows spiral-like behaviour. Reconsidering our results of Hopf bifurcation involved in the Eocene-Oligocene transition, these spirals might represent relaxation onto a shifting bound state in the Eocene. A similar picture is found for the optimal delay embedding $\tau = 25$ in subfigure \ref{subfig:  nd tau25}. The attractor is significantly better unfolded and fills large parts of the state space. However, it is not randomly spread, but confined to a certain region, which indicate non-random dynamics. Furthermore, we observe again spiral-like shrinking dynamics over time. Using again higher delay time as $\tau = 140$ in subfigure \ref{subfig:  nd tau140} also provides an unfolded trajectory which is bound in state space, but the chosen delay time might introduce a significant irrelevance error~\cite{a:rosenstein} and leads to a very sparse data base.\\
For completeness we present the state space visualization also for the detrended data with a Gaussian kernelwidth $\tilde{\sigma}=5$ for three different delay times $\tau$. The delays are chosen to be comparable in terms of the add results (for the add results cf. subfigures \ref{subfig: time lag} and \ref{subfig: time lag detrended}). The distribution of the residuals in state space confirms that the detrending cuts off the spiral-like dynamics and the drift of the assumed bound state. In such a representation, the residual data is successively stronger confined to the shifted attracting region around $(0,0,0)$ in all cases. This is well observable from the outer blue to the inner red data points.\\
In figure \ref{fig: shadow attractors whole} we provide an analogue analysis of the whole $CaCO_3$ time series from the Pacific sediment core at site DSDP1218. The add results for the whole time series are summarized in figure \ref{fig: delay whole set}. Similar to our previous analysis we find high redundancy error for small time delays and probably higher irrelevance error and significantly less data for high time delays. Nevertheless, the trajectory is non-randomly spread. Compared to the analysis in figure \ref{fig: shadow attractors analysed period} the most interesting difference comes from the transient data in the beginning and the end of the considered whole time interval. The blue transient data in the beginning of the time series travel into the more bound region of the time interval from $\SIrange{34.03}{39.93}{mya}$ (light blue to yellow). This might reflect the transient dynamics of the reported deepening in
calcite compensation depth $\sim\SI{42}{mya}$ into a more confined Eocene climate state~\cite{a:Tripati2005, d:Tripati2005}. The data $<\SI{34.03}{mya}$ (yellow to red) exhibit a well observable spiral-like behaviour and significant drift culminating in the red branch that seems to escape from the previous bound region. These observations fit into our findings of a Hopf bifurcation involved in the Eocene-Oligocene transition, since the spiral-like behaviour implies periodic features. However, the escaping red branch clearly shows that the overall picture is much more complicated and needs further research to understand it better. Finally, we show three embeddings of the whole time series detrended with a Gaussian kernelwidth $\tilde{\sigma}=20$ in figure \ref{fig: shadow attractors whole}(b, d, f). Especially the final drift of the orange to red data points is not cancelled by the Gaussian detrending with bandwidth $\tilde{\sigma} = 20$. Even a very strong detrending approach with kernelwidth $\tilde{\sigma} = 5$ does not cancel this deviating orange-red branch in the uprising Oligocene completely which might underline the strength of the shifting dynamics.\\

\subsection{Robustness of the bifurcation analysis}
\label{subsec: appendix greenhouse result robustness}
\begin{figure*}
    \centering
    \includegraphics[width=0.85\textwidth]{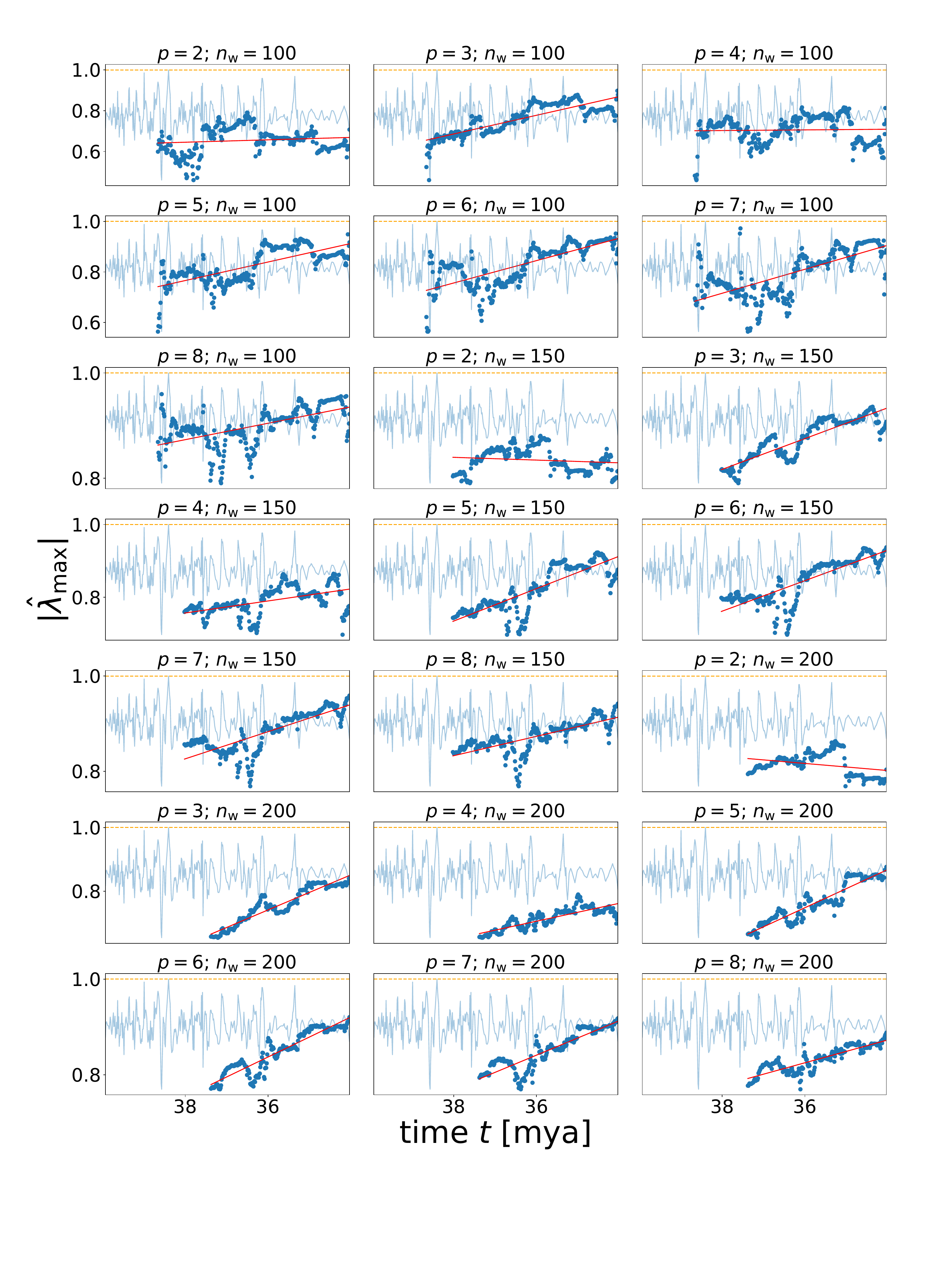}
    \caption{Robustness check of the absolute DEV $|\hat{\lambda}_{\rm max}|$ analysis of the $\delta(CaCO_3)$ concentration residuals for a Gaussian detrending with kernelwidth $\tilde{\sigma} = 5$. The results are robust against changes in window size $n_{\rm w}$ and embedding dimension, i.e. AR orders $p$. Only an embbeding in two dimensions which lies below the minimum proper dimension of $3$ proposed by the fnn algorithm does not exhibit clear positive trends as well as the result for window size $n_{\rm w} = 100$ with $p=4$.}
    \label{fig: robustness DEV detrend5}
\end{figure*}

\begin{figure*}
    \centering
    \includegraphics[width=0.85\textwidth]{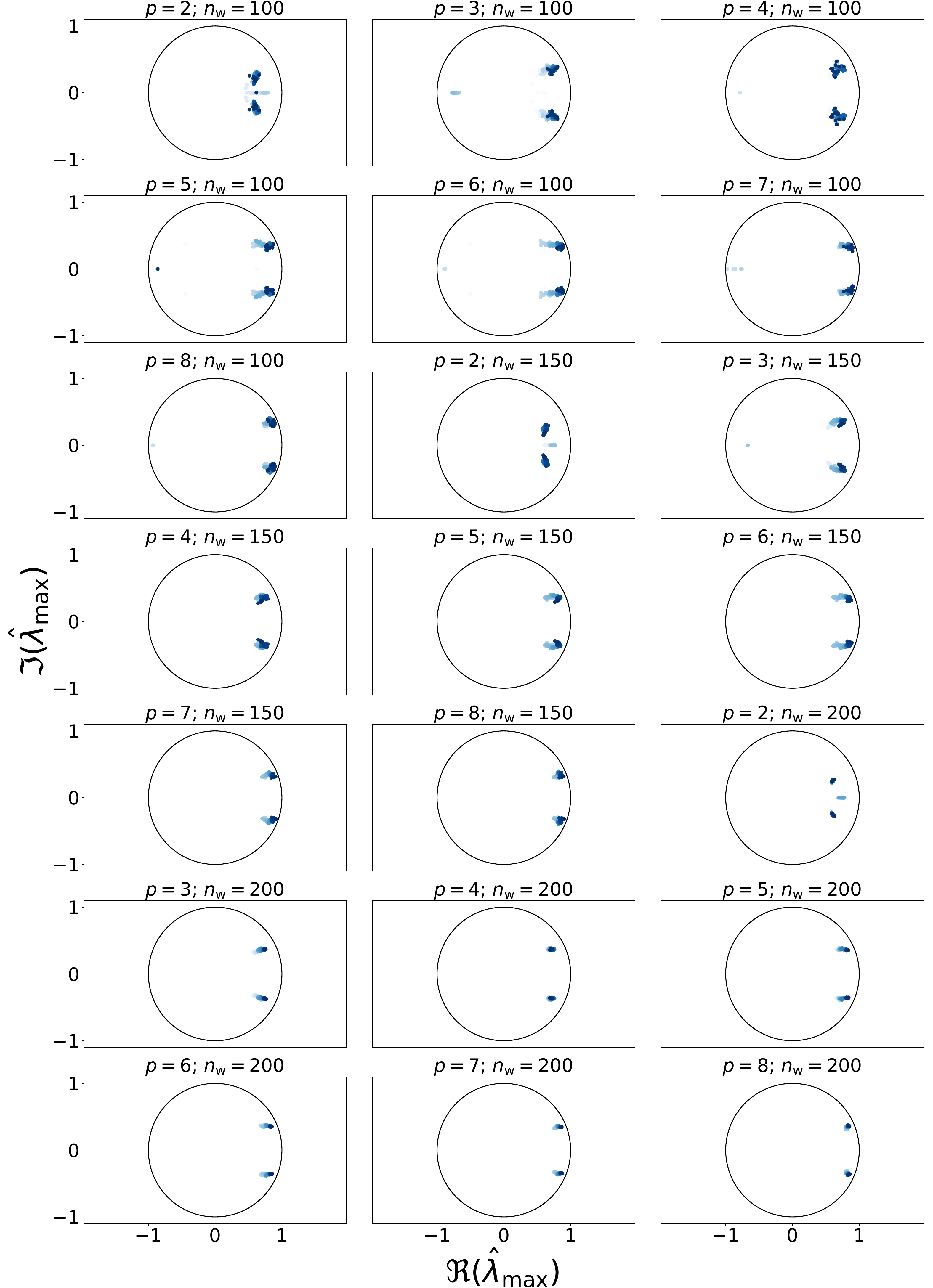}
    \caption{Same case as in figure \ref{fig: robustness DEV detrend5} for the DEV $\hat{\lambda}_{\rm max}$ bifurcation classification in the Gaussian plane. Time evolution is resolved from transparent to opaque points. Regardless of window sizes $n_{\rm w}$ and AR model orders $p$ the Neimark-Sacker bifurcation, which implies uprising periodicity, is suggested by our analysis. Some numerical artefacts arise from a too low embedding dimension $p=2$ with some pure real negative DEV $\hat{\lambda}_{\rm max}$. Additionally, the DEV estimation in the Gaussian plane is slightly more affected by decreasing data availability as can be observed by similar numerical artefacts for $n_{\rm w}=100$.}
    \label{fig: robustness imagDEV detrend5}
\end{figure*}

\begin{figure*}
    \centering
    \includegraphics[width=0.85\textwidth]{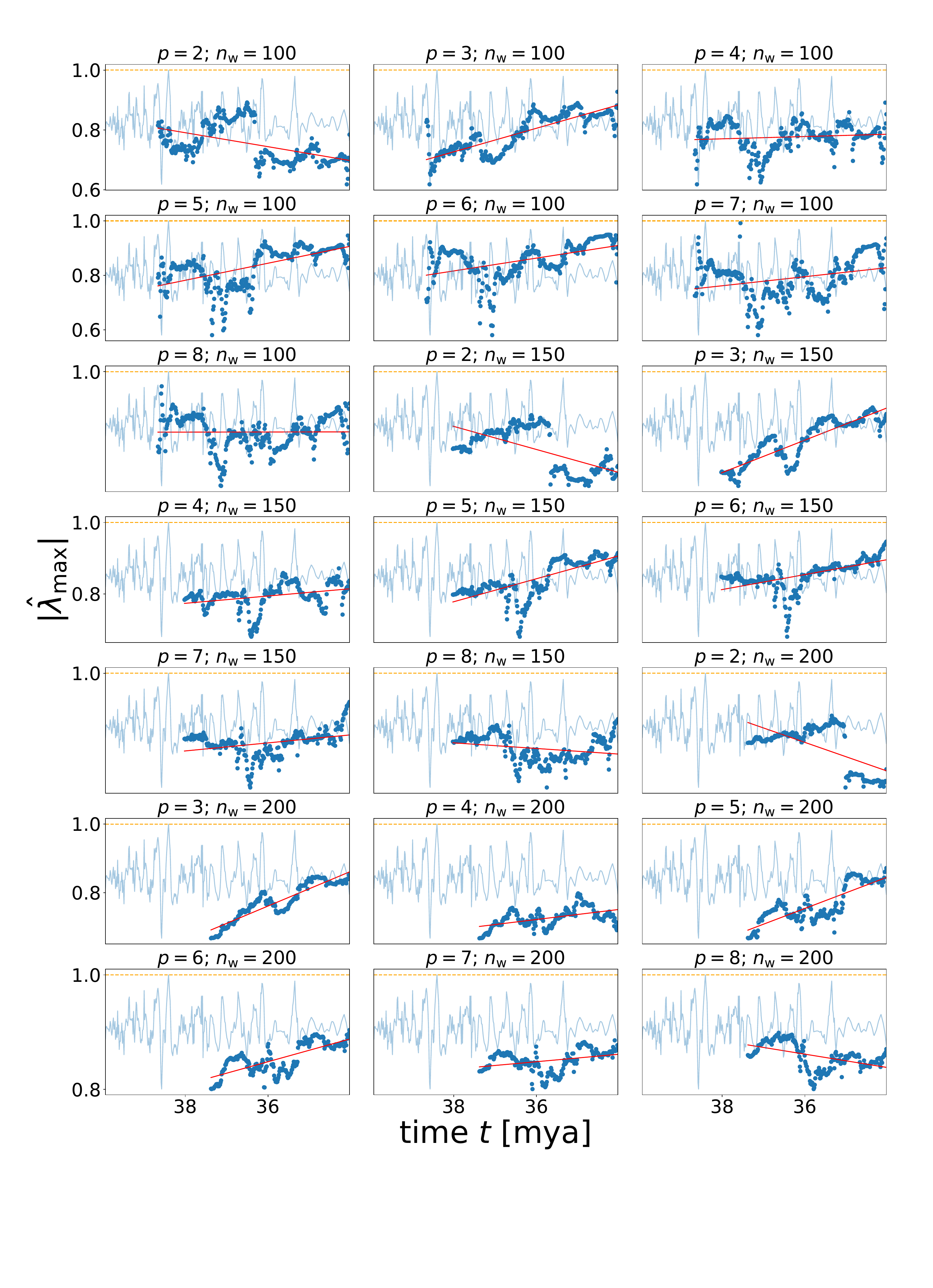}
    \caption{Robustness check of the absolute DEV $|\hat{\lambda}_{\rm max}|$ analysis of the $\delta(CaCO_3)$ concentration residuals for a Gaussian detrending with kernelwidth $\tilde{\sigma} = 10$. The qualitative results are almost identical to figure \ref{fig: robustness DEV detrend5}.}
    \label{fig: robustness DEV detrend10}
\end{figure*}

\begin{figure*}
    \centering
    \includegraphics[width=0.85\textwidth]{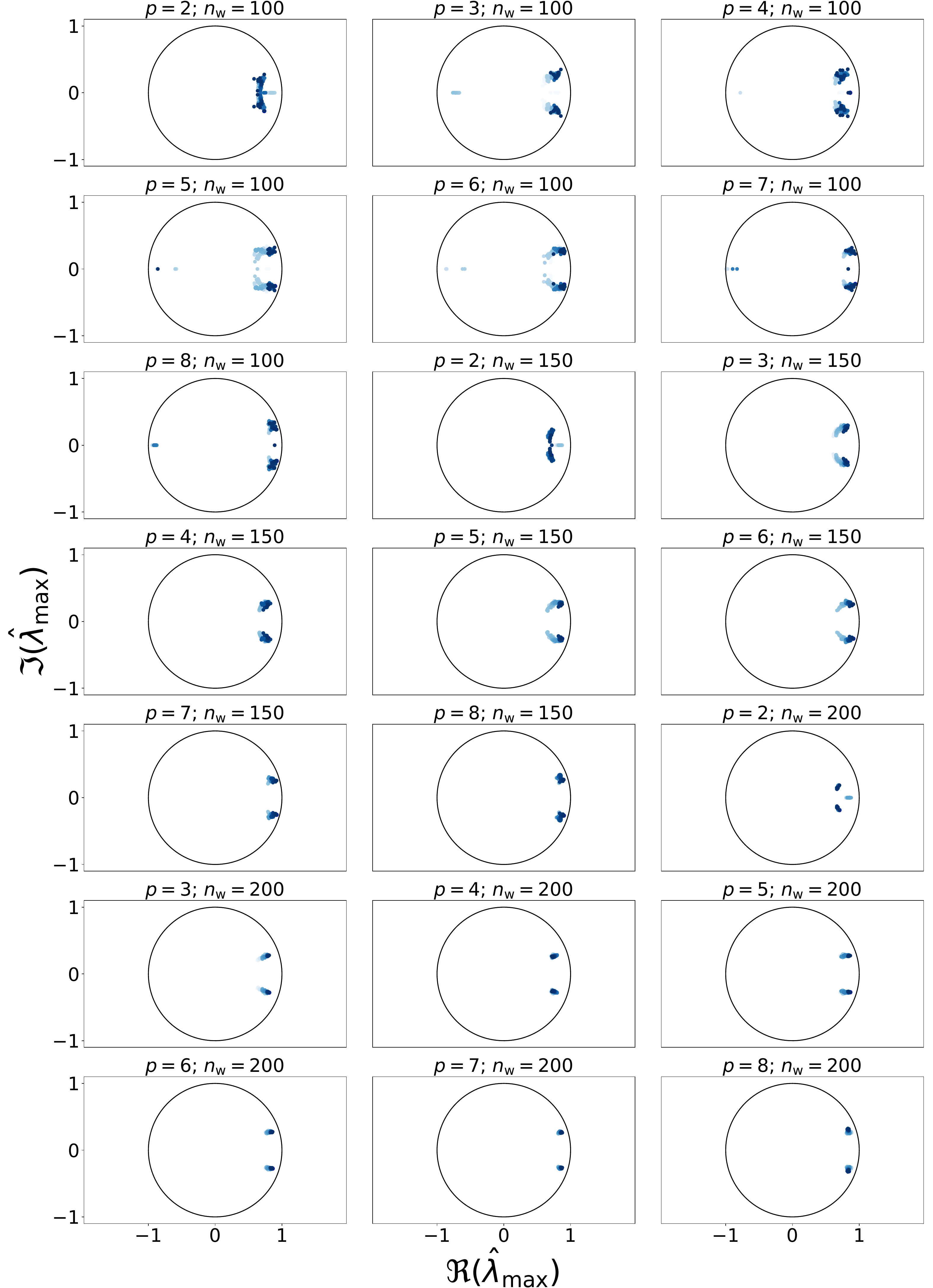}
    \caption{Same case as in figure \ref{fig: robustness DEV detrend10} for the DEV $\hat{\lambda}_{\rm max}$ bifurcation classification in the Gaussian plane. Time evolution is resolved from transparent to opaque points. The results for the detrending bandwidth $\tilde{\sigma}= 10$ are qualitatively almost identical to the findings for the detrending bandwidth bandwidth $\tilde{\sigma}= 5$ (cf. figure \ref{fig: robustness DEV detrend5}).}
    \label{fig: robustness imagDEV detrend10}
\end{figure*}

\begin{figure*}
    \centering
    \includegraphics[width=0.85\textwidth]{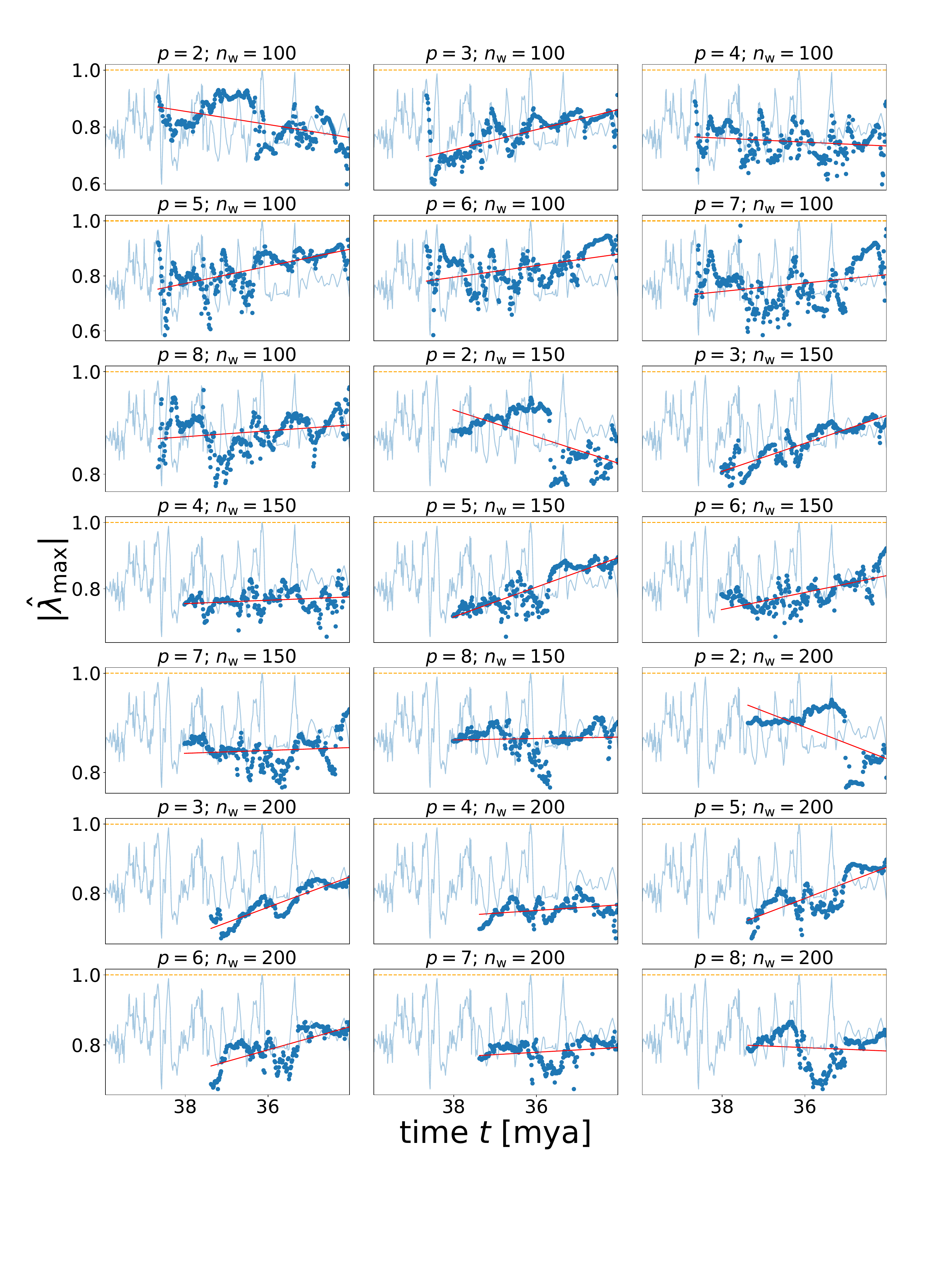}
    \caption{Robustness check of the absolute DEV $|\hat{\lambda}_{\rm max}|$ analysis of the $\delta(CaCO_3)$ concentration residuals for a Gaussian detrending with kernelwidth $\tilde{\sigma} = 20$.  The qualitative results are almost identical to the figures \ref{fig: robustness DEV detrend5} and \ref{fig: robustness DEV detrend10}.}
    \label{fig: robustness DEV detrend20}
\end{figure*}

\begin{figure*}
    \centering
    \includegraphics[width=0.75\textwidth]{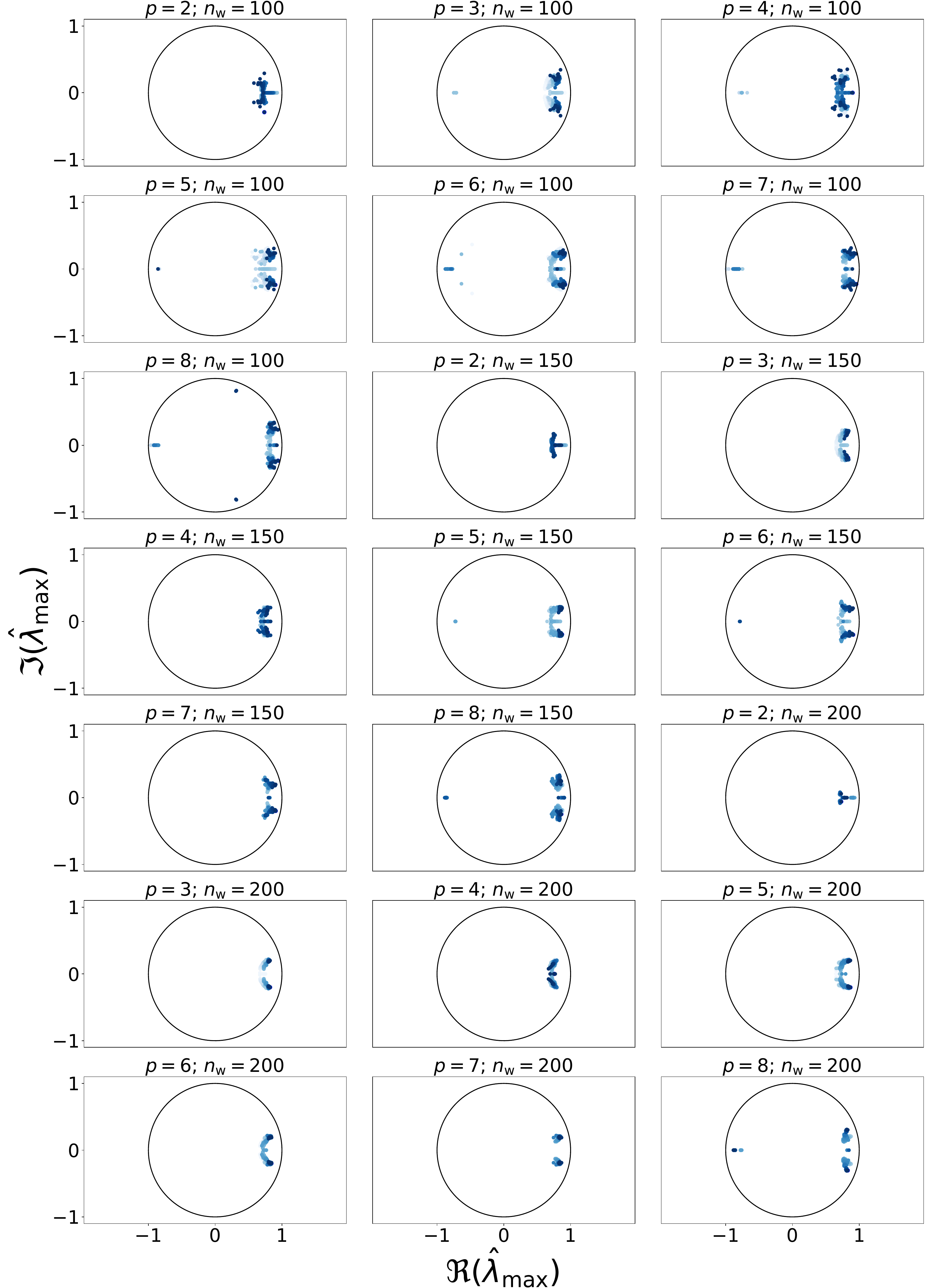}
    \caption{
    Same case as in figure \ref{fig: robustness DEV detrend20} for the bifurcation classification. Weaker detrending (bandwidth $\tilde{\sigma}=20$) has the greatest effect on our results compared to other parameters. This is an expected result. A weaker detrending probably leads to improper subtraction of non-stationary trends. Since the Jacobian parameters of the DEV analysis are estimated via AR($p$) which holds only for stationary data, the estimates become more and more perturbed due to partially preserved slow non-stationary effects. The results for window size $n_{\rm w} = 100$ scatter heavily. Windows sizes $n_{\rm w}\gtrsim 150$ tend to stabilize the estimation of complex conjugated eigenvalues that suggest rising periodic dynamics by a Neimark-Sacker bifurcation. In most of the cases, the estimates early in time scatter on the real axis. However, the estimates become typically complex conjugates with modulus approaching unity when time goes on. As observed for other bandwidths the estimates for order $p=2$ are not trustworthy, since they are smaller than the optimal range of embedding dimensions, i.e. roughly $d_{\rm opt} = \SIrange[]{3}{5}{}$.}
    \label{fig: robustness imagDEV detrend20}
\end{figure*}

In this subsection we perform a robustness check of our results regarding the choice of the Gaussian kernel bandwidth $\tilde{\sigma}$, the embedding dimension $d$ which is equivalent to order $p$ of the AR scheme in our approach and the rolling window size $n_{\rm w}$. We repeated the analysis for all combinations of the following parameters: 
\begin{itemize}
    \item window sizes $n_{\rm w} = \lbrace 100,~150,~200\rbrace$,
    \item embedding dimensions $d_{\rm opt} = p = \lbrace 2,~3,~4,~5,~6,~7,~8\rbrace$,
    \item and Gaussian kernel bandwidths $\hat{\sigma} = \lbrace 5,~10,~20\rbrace$ for the detrending procedure.
\end{itemize}
The results are presented in the figures \ref{fig: robustness DEV detrend5}, \ref{fig: robustness DEV detrend10} and \ref{fig: robustness DEV detrend20}. If the optimal embedding range $d_{\rm opt}\approx\SIrange[]{3}{5}{}$ determined by the fnn algorithm is chosen, the positive absolute DEV trend is reproduced for all window sizes $n_{\rm w}$ and kernelwidths with exception of window size $n_{\rm w} = 100$ with $p=4$. A smaller embedding dimension $d<3$ leads generally to negative trends, whereas the positive trends are damped and sometimes reversed for higher embedding dimension $d>5$.\\ 
A visualization of the DEV estimates for the various parameter combinations and detrending bandwidths is shown in the figures \ref{fig: robustness imagDEV detrend5}, \ref{fig: robustness imagDEV detrend10} and \ref{fig: robustness imagDEV detrend20}. The DEV results in the Gaussian plane stabilize for windows sizes of $n_{\rm w} \geq 150$ for the kernelwidth $\tilde{\sigma}=\lbrace 5,~10\rbrace$. For $n_{\rm w}=100$ there are some pure real DEV estimates relatively early in time. The DEV estimates in the Gaussian plane are mostly disturbed by weaker detrending as seen for the kernelwidth $\tilde{\sigma}=20$. This is an expected result, since the AR($p$) estimation is only applicable to stationary time series. A kernelwidth $\tilde{\sigma}=20$ might lead to imperfect subtraction of the slow non-stationary trend which leads to disturbed DEV estimates. In the limit of no detrending at all the AR$(p)$ based DEV estimation does not provide any meaningful estimates anymore. The values scatter in the Gaussian plane, sometimes pure real, sometimes complex conjugated and they often exhibit negative trends in the modulus. In that sense, the S-map based DEV approach proves to be slightly more stable against improper detrending. However, also in the S-map DEV estimates of Grziwotz et al.~\cite{a:grziwotz} the DEVs scatter around the real axis with slightly non-zero imaginary parts.
\end{document}